\documentclass{article}

\usepackage{PRIMEarxiv}

\usepackage{tikz}
\usepackage{float}
\usepackage{comment}
\usepackage{tabularx}
\usepackage{subfigure}
\usepackage{array, makecell} 
\usepackage{arydshln}
\usepackage{pifont}
\usepackage{tablefootnote}
\usepackage{fontenc, inputenc, calc, indentfirst, fancyhdr, graphicx, epstopdf, lastpage, ifthen, lineno, float, amsmath, setspace, enumitem, mathpazo, booktabs, titlesec, etoolbox, tabto, xcolor, soul, multirow, microtype, tikz, totcount, changepage, attrib, upgreek, amsthm, hyphenat, footmisc, url, geometry, newfloat, caption}
\usepackage[sort&compress,numbers]{natbib}

\definecolor{bluecite}{HTML}{0875b7}
\usepackage[colorlinks=true,linkcolor=bluecite,citecolor=bluecite,urlcolor=bluecite]{hyperref}

\newcommand*\rot{\rotatebox{90}}
\newcommand*\OK{\ding{51}}

\titleformat{\paragraph} {\raggedright\fontsize{10}{10}\selectfont}{\theparagraph.\space}{0pt}{}
\titlespacing{\paragraph} {0pt} {12pt} {3pt}

\newlength{\extralength}
\newcolumntype{P}[1]{>{\centering\arraybackslash}p{#1}}
\newcolumntype{C}{>{\centering\arraybackslash}X} 
  
\title{Self-Sovereign Identity: A Systematic Review, Mapping and Taxonomy}

\author{
  Frederico Schardong \\
  Federal Institute of Education, Science and Technology of Rio Grande do Sul \\
  Rolante, Brazil \\
  Federal University of Santa Catarina \\
  Florianópolis, Brazil \\
  \texttt{frederico.schardong@rolante.ifrs.edu.br} \\
   \And
  Ricardo Custódio \\
  Federal University of Santa Catarina \\
  Florianópolis, Brazil \\
  \texttt{ricardo.custodio@ufsc.br} \\
}

\begin{document}
\maketitle

\begin{abstract}
Self-Sovereign Identity (SSI) is an identity model centered on the user. The user maintains and controls their data in this model. When a service provider requests data from the user, the user sends it directly to the service provider, bypassing third-party intermediaries. Thus, SSI reduces identity providers' involvement in the identification, authentication, and authorization, thereby increasing user privacy. Additionally, users can share portions of their personal information with service providers, significantly improving user privacy. This identity model has drawn the attention of researchers and organizations worldwide, resulting in an increase in both scientific and non-scientific literature on the subject. This study conducts a comprehensive and rigorous systematic review of the literature and a systematic mapping of theoretical and practical advances in SSI. We identified and analyzed evidence from reviewed materials to address four research questions, resulting in a novel SSI taxonomy used to categorize and review publications. Additionally, open challenges are discussed, along with recommendations for future work.
\end{abstract}




\section{Introduction}\label{sec:introduction}
The ability to prove that individuals are who they claim to be is critical to human interactions in society, whether in the physical world or online. The proof is typically presented in the form of a credential that enables the identification and authentication of a person. This credential, which consists of a collection of attributes, is referred to as an identity document or simply identity~\cite{ISO24760,bertino2010identity}.

In today's digital world, large corporations such as Google and Facebook issue electronic identities. They created these identities to facilitate user identification, authentication, authorization, and provision of user attributes for their internal services. These identities have developed into a powerful tool for identifying users who wish to access the companies' services and those of a variety of other Service Providers (SPs). As a result, these businesses serve as Identity Providers (IdPs). Numerous companies have outsourced their customer registration, identification, and authentication to IdPs.

Using IdPs has a number of benefits and drawbacks. The user benefits from having a single identity to authenticate with multiple SPs. One disadvantage may be that a single IdP manages data for many users. Storing people's electronic identities in a few IdPs has been a source of contention due to the fact that these few data silos have the data of a large number of people~\cite{Path2016}. These massive data silos have become attractive targets for hackers~\cite{linkedin2016} because they contain high-value assets that can be misused~\cite{isaak2018user} or even traded~\cite{fastcompany2019} with institutions that users have not authorized.

Although the vast majority of users trust IdPs naively, many users and businesses are uneasy with the requirement to use and trust these entities. Self-Sovereign Identity (SSI)~\cite{Path2016} has garnered attention in this context because it prevents IdPs from tracking their users' activities. Additionally, it also enhances people's privacy by enabling them to store and manage their data and specify the granularity of the information they can share.

Despite the fact that SSI provides sovereignty over the digital presence, it introduces new challenges that must be overcome before widespread adoption can occur. The difficulties are conceptual and pragmatic in nature. The primary conceptual problems are defining SSI and defining what constitutes a self-sovereign system. The pragmatic challenges include, but are not limited to, how to coexist with and migrate existing IdPs' identities to the new model, how to trust data from other self-sovereign identities, and how to assist users with managing, backing up, and recovering private data.

The advantages of this new identity paradigm over traditional models have attracted researchers' and professionals' attention in recent years, resulting in an increasing number of publications on the subject. Some initiatives aim to review and condense the body of knowledge thus far. However, current reviews do not address all facets of SSI. For instance, they omit publications that contribute to the conceptual debate over the meaning of the term ``self-sovereign identity'' and efforts that present novel problems and solutions in specific areas of SSI. Existing reviews are primarily concerned with applications and research papers that propose SSI systems such as Sovrin~\cite{sovrin} and uPort~\cite{uport}. 

This article conducts a comprehensive systematic review and mapping of the scientific and non-scientific literature that contribute to the debate over what SSI is, as well as works that address practical issues related to SSI. We searched for, selected, and reviewed publications in a systematic manner, guided by four research questions. Due to the systematic nature of our work, it may be reproduced and updated in the future to reflect new activity. The results include: (i) a taxonomy that enables hierarchical classification of the SSI literature; (ii) an in-depth and systematic analysis of the surveyed materials using our novel taxonomy; and (iii) analyses and maps of publication frequency, venues, co-references, and co-authorships, which provide a global view of the state of the art of SSI literature to the reader. Finally, open issues and recommendations for researchers and practitioners working with SSI are discussed.

\subsection{Novelty and Research Contributions}

In summary, we make the following three main research contributions to the field.

\begin{itemize}
    \item Our survey examines both \emph{conceptual} and \emph{practical} advances in SSI, highlighting philosophical contributions to the definition of SSI, novel problems and proposed solutions, and promising directions for \emph{future research}. The manuscript conducts an analysis of the body of knowledge established by over 80 research papers, scientific reports, patents, technological standards, and theses.
    
    \item Through a proposed \emph{taxonomy}, we provide the reader with a comprehensive and organized understanding of the SSI literature. Additionally, the manuscript presents and discusses \emph{maps} of authors' relationships, publication venues, and the shift in the focus of research in the area over time. To our knowledge, this is the first survey of SSI to include a \emph{systematic literature review}, a \emph{systematic mapping}, and a \emph{taxonomy}, all of which are based on \emph{rigorous} criteria and \emph{reproducible methodology}.
    
    \item Unlike previous surveys~\cite{Liu2020b,vcuvcko2021decentralized,ghaffari2021identity,mulaji2021practicality,Kuperberg2019,zhu2018identity,Lim2018,kaneriya2020comparative,gilani2020survey,dib2020decentralized,Muhle}, we examine \emph{conceptual} discussions of SSI and include publications that are not \emph{blockchain-based}.
\end{itemize}

The remainder of this article is structured in the following manner. Section~\ref{sec:background} provides an introduction to electronic identity and a detailed description of SSI. Section~\ref{sec:related} outlines the existing secondary studies that review the SSI literature and their shortcomings. Section~\ref{sec:method} defines the method used in this study and how it was carried out. Section~\ref{sec:taxonomy} presents the reader with the proposed taxonomy. In Section~\ref{sec:RQ1}, we describe the practical research surveyed. Section~\ref{sec:RQ2} identifies and discusses mathematical and cryptographic tools used in applied research. In Section~\ref{sec:RQ3}, we detail philosophical discussions regarding understanding what SSI is, and in Section~\ref{sec:RQ4}, we present the results of our mapping. Finally, in Section~\ref{sec:challenges}, we discuss the open challenges and shortcomings, and in Section~\ref{sec:conclusion}, we make final remarks.

\section{Preliminaries}\label{sec:background}
This section provides the necessary background for following this study. We begin with an introduction to identity documents and then discuss electronic identities and their evolution into SSI, which we describe in detail.

\subsection{Identity Documents}

We can categorize identity documents into three distinct formats of representation. The traditional physical document is the first format. This format typically consists of a paper document or a plastic card on which the individual's identifying characteristics are printed. Paper and plastic cards are manufactured with care to avoid easy forgery. When a person wishes to prove their identity, they present a physical document. The Relying Party (RP) performs the identification by reading the attributes. One of the most critical characteristics of this type of document is the photograph of the individual's face, which is used for authentication. This identification document is referred to as a face badge.

The digital identity document is the second format. It can be thought of as the digital version of the physical document and is often used on mobile devices~\cite{id123}. Cryptographic techniques, such as digital signatures, are used to verify the integrity and authenticity of the data. Typically, the signature and identity attributes are encoded as a QR code so that the RP can verify the identity document's integrity and veracity offline.

The electronic identity document is the third format. This is the identity that is used in the virtual world to authenticate users and enable them to consume electronic services on the web. Unlike a digital document, which is a visual representation, an electronic document is built from the ground up to be used electronically, removing the need for visual verification of its integrity. Multi-factor authentication~\cite{cryptography2010001} and cryptographic techniques such as digital signatures and public-key cryptography~\cite{Stallings} are used to carry out these processes. For instance, by combining a password known only to the identity holder with a key displayed in a time-based one-time password service~\cite{rfc6238,erdem2018otpaas}.

These three forms of identity must be impervious to forgery, fraud, and data leakage. As a result, the collection, storage, and processing of identity-related data must be handled with extreme caution, with an emphasis on the use of appropriate data protection mechanisms. While each of the three types of identity listed above is vulnerable to fraud, the electronic version requires the most oversight. Numerous instances of fraud involving the misuse of electronic identities have been reported~\cite{isaak2018user,fastcompany2019}.

While business cards and \textit{curriculum vitae} are examples of self-issued identity documents, the vast majority of identities in use are issued by trusted third parties. For instance, national-level identification documents such as driver's licenses 
and passports are frequently issued by the government~\cite{world2019id4d} or by private companies authorized to do so~\cite{govuk}.

\subsection{Electronic Identity}

In the physical world, establishing trust in relationships between various entities requires identifying the communicating parts. Proof of identity is accomplished through pre-agreed upon authentication factors or with the assistance of trusted third parties. Physical devices are frequently used as authentication factors. For instance, it is not uncommon for individuals to be identified visually through their identification documents, followed by a facial badge verification. Similarly, in the electronic world, communicating parties must have a certain Level of Assurance (LoA) regarding the other party's identity. This assurance is accomplished through the use of electronic identities on data communication networks such as the Internet.

As with a physical identity, an electronic identity is typically defined as a set of attributes that help in the description or qualification of an entity~\cite{ISO24760}. Some authors prefer to limit this definition to a set of attributes in a specific context in order to improve its accuracy~\cite{miyata2006survey, el2007survey, ferdous2015user}. As a result, electronic identities are not simply digital representations of physical identities such as a passport or driver's license. They are created, used, and destroyed in accordance with the user's desires, frequently containing only the attributes necessary to accomplish the task at hand. For instance, a seller on eBay~\cite{ebay} may have an electronic identity that conceals their name, age, and country of residence, as others are only concerned about whether or not this seller has a track record of successful transactions~\cite{resnick2006value}.

All identities, whether physical or electronic, are subject to ownership verification. That is, they require mechanisms for properly \emph{identifying} and \emph{authenticating} users~\cite{kiennert2015authentication}.

The identification process begins with the holder of an electronic identity presenting a unique attribute in a given context, \textit{i.e.}, an identifier that differentiates it from all other electronic identities in that context~\cite{ferdous2014mathematical}. The most common example is providing an email address when signing up for a subscription service. The subsequent stage is to authenticate the identified entity by verifying a security proof, which is traditionally accomplished via a secret password or digital signature, thereby ensuring that the holder of the identity is, in fact, its owner. In the above-mentioned subscription service example, providing a code or clicking a link received via email proves that the email address belongs to the holder.

Identification and authentication are critical in our digital society because they enable citizens to access services electronically. As a result, the identification and authentication processes are carried out by specialized services trusted by the parties involved. These services are provided by systems that manage electronic identity and are referred to as Identity and Access Management (IAM) systems.

\subsection{The evolution of IAM models}

In the early days of the web, SPs had to implement their own IAM solutions to identify and authenticate clients to offer personalized products and services. As a result, these services are referred to as \emph{centralized authorities}. This model presented a number of usability issues for users. Most users ended up using similar low-entropy passwords on different systems, making room for numerous vulnerabilities. This model has sparked numerous initiatives aimed at educating users about the dangers of using simple passwords and reusing them across multiple services~\cite{talib2010analysis, scarfone2009guide}.

The next logical evolution was to replace the centralized authorities with third-party IAM solutions, \textit{i.e.}, IdPs. With this new paradigm, users only need to be registered with a few IdPs in order to access the web's plethora of services. By contrast, SPs must be registered with the desired IdPs or IdP federations to work with the IdPs' identified and authenticated users. Through token exchange protocols such as SAML~\cite{hughes2005security}, OAuth 2.0~\cite{hardt2012oauth}, and OpenID Connect~\cite{recordon2006openid}, the interactions between the IdP, SP, and end user were standardized. Even though this identity model significantly simplified the management of multiple identifiers and passwords for users, it resulted in the creation of a few large silos of valuable private information.

The user-centric model was the next evolutionary step~\cite{josang2005user}. It was designed with the idea that users could use Personal Authentication Devices (PADs), such as smartphones and smartcards, to store and present authentication credentials from SPs, bypassing the need for third-party IdPs. However, as noted in~\cite{Path2016}, this model has not gained traction and is currently viewed as an extension of the IdP model with greater user control. According to~\cite{Path2016}, the current interpretation of this model is that the user is aware of and must authorize or deny her IdP sharing specific personal attributes requested by an SP. As a result, the current model of user-centric identity faces the same issues as the previous model.

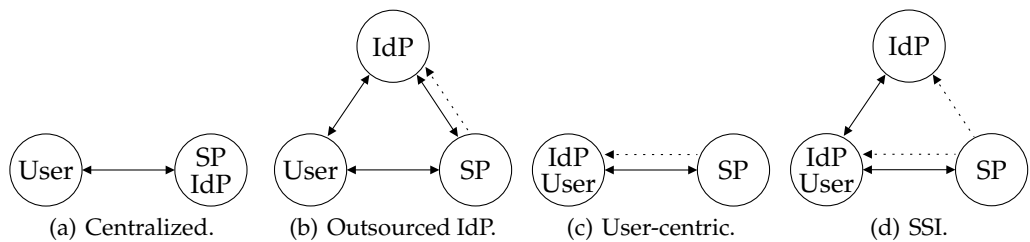
\begin{figure}[H]
    \centering
    \subfigure[Centralized.]  
{  
    \begin{tikzpicture}[x=0.39pt,y=0.39pt,yscale=-1,xscale=1]
        \draw   (20,1715) .. controls (20,1695.67) and (35.67,1680) .. (55,1680) .. controls (74.33,1680) and (90,1695.67) .. (90,1715) .. controls (90,1734.33) and (74.33,1750) .. (55,1750) .. controls (35.67,1750) and (20,1734.33) .. (20,1715) -- cycle ;
        \draw   (180,1715) .. controls (180,1695.67) and (195.67,1680) .. (215,1680) .. controls (234.33,1680) and (250,1695.67) .. (250,1715) .. controls (250,1734.33) and (234.33,1750) .. (215,1750) .. controls (195.67,1750) and (180,1734.33) .. (180,1715) -- cycle ;
        \draw    (93,1715) -- (177,1715) ;
        \draw [shift={(180,1715)}, rotate = 540] [fill={rgb, 255:red, 0; green, 0; blue, 0 }  ][line width=0.08]  [draw opacity=0] (8.93,-4.29) -- (0,0) -- (8.93,4.29) -- cycle    ;
        \draw [shift={(90,1715)}, rotate = 360] [fill={rgb, 255:red, 0; green, 0; blue, 0 }  ][line width=0.08]  [draw opacity=0] (8.93,-4.29) -- (0,0) -- (8.93,4.29) -- cycle    ;
        
        \draw (55,1715) node   [align=left] {User};
        \draw (215,1702) node   [align=left] {SP};
        \draw (215,1728) node   [align=left] {IdP};
    \end{tikzpicture}
    \label{fig:iam:centralized}
}  
\subfigure[Outsourced IdP.]  
{  
    \begin{tikzpicture}[x=0.39pt,y=0.39pt,yscale=-1,xscale=1]
        \draw   (20,1945) .. controls (20,1925.67) and (35.67,1910) .. (55,1910) .. controls (74.33,1910) and (90,1925.67) .. (90,1945) .. controls (90,1964.33) and (74.33,1980) .. (55,1980) .. controls (35.67,1980) and (20,1964.33) .. (20,1945) -- cycle ;
        \draw   (180,1945) .. controls (180,1925.67) and (195.67,1910) .. (215,1910) .. controls (234.33,1910) and (250,1925.67) .. (250,1945) .. controls (250,1964.33) and (234.33,1980) .. (215,1980) .. controls (195.67,1980) and (180,1964.33) .. (180,1945) -- cycle ;
        \draw   (100,1825) .. controls (100,1805.67) and (115.67,1790) .. (135,1790) .. controls (154.33,1790) and (170,1805.67) .. (170,1825) .. controls (170,1844.33) and (154.33,1860) .. (135,1860) .. controls (115.67,1860) and (100,1844.33) .. (100,1825) -- cycle ;
        \draw    (110.32,1855.08) -- (73.28,1909.72) ;
        \draw [shift={(71.6,1912.2)}, rotate = 304.13] [fill={rgb, 255:red, 0; green, 0; blue, 0 }  ][line width=0.08]  [draw opacity=0] (8.93,-4.29) -- (0,0) -- (8.93,4.29) -- cycle    ;
        \draw [shift={(112,1852.6)}, rotate = 124.13] [fill={rgb, 255:red, 0; green, 0; blue, 0 }  ][line width=0.08]  [draw opacity=0] (8.93,-4.29) -- (0,0) -- (8.93,4.29) -- cycle    ;
        \draw    (159.66,1855.1) -- (196.34,1910.5) ;
        \draw [shift={(198,1913)}, rotate = 236.49] [fill={rgb, 255:red, 0; green, 0; blue, 0 }  ][line width=0.08]  [draw opacity=0] (8.93,-4.29) -- (0,0) -- (8.93,4.29) -- cycle    ;
        \draw [shift={(158,1852.6)}, rotate = 56.49] [fill={rgb, 255:red, 0; green, 0; blue, 0 }  ][line width=0.08]  [draw opacity=0] (8.93,-4.29) -- (0,0) -- (8.93,4.29) -- cycle    ;
        \draw    (177,1945) -- (93,1945) ;
        \draw [shift={(90,1945)}, rotate = 360] [fill={rgb, 255:red, 0; green, 0; blue, 0 }  ][line width=0.08]  [draw opacity=0] (8.93,-4.29) -- (0,0) -- (8.93,4.29) -- cycle    ;
        \draw [shift={(180,1945)}, rotate = 180] [fill={rgb, 255:red, 0; green, 0; blue, 0 }  ][line width=0.08]  [draw opacity=0] (8.93,-4.29) -- (0,0) -- (8.93,4.29) -- cycle    ;
        \draw  [dash pattern={on 0.84pt off 2.51pt}]  (168.07,1847.49) -- (210,1910) ;
        \draw [shift={(166.4,1845)}, rotate = 56.15] [fill={rgb, 255:red, 0; green, 0; blue, 0 }  ][line width=0.08]  [draw opacity=0] (8.93,-4.29) -- (0,0) -- (8.93,4.29) -- cycle    ;

        \draw (55,1945) node   [align=left] {User};
        \draw (215,1945) node   [align=left] {SP};
        \draw (135,1825) node   [align=left] {IdP};
    \end{tikzpicture}
    \label{fig:iam:outsourced}
}
\subfigure[User-centric.]{
    \begin{tikzpicture}[x=0.39pt,y=0.39pt,yscale=-1,xscale=1]
        \draw   (20,2085) .. controls (20,2065.67) and (35.67,2050) .. (55,2050) .. controls (74.33,2050) and (90,2065.67) .. (90,2085) .. controls (90,2104.33) and (74.33,2120) .. (55,2120) .. controls (35.67,2120) and (20,2104.33) .. (20,2085) -- cycle ;
        \draw   (180,2085) .. controls (180,2065.67) and (195.67,2050) .. (215,2050) .. controls (234.33,2050) and (250,2065.67) .. (250,2085) .. controls (250,2104.33) and (234.33,2120) .. (215,2120) .. controls (195.67,2120) and (180,2104.33) .. (180,2085) -- cycle ;
        \draw    (93,2085) -- (177,2085) ;
        \draw [shift={(180,2085)}, rotate = 180] [fill={rgb, 255:red, 0; green, 0; blue, 0 }  ][line width=0.08]  [draw opacity=0] (8.93,-4.29) -- (0,0) -- (8.93,4.29) -- cycle    ;
        \draw [shift={(90,2085)}, rotate = 0] [fill={rgb, 255:red, 0; green, 0; blue, 0 }  ][line width=0.08]  [draw opacity=0] (8.93,-4.29) -- (0,0) -- (8.93,4.29) -- cycle    ;
        \draw  [dash pattern={on 0.84pt off 2.51pt}]  (93,2070) -- (180,2070) ;
        \draw [shift={(90,2070)}, rotate = 0] [fill={rgb, 255:red, 0; green, 0; blue, 0 }  ][line width=0.08]  [draw opacity=0] (8.93,-4.29) -- (0,0) -- (8.93,4.29) -- cycle    ;
    
        \draw (55,2098) node   [align=left] {User};
        \draw (215,2085) node   [align=left] {SP};
        \draw (55,2072) node   [align=left] {IdP};
    \end{tikzpicture}
    \label{fig:iam:usercentric}
}
\subfigure[SSI.]{
    \begin{tikzpicture}[x=0.39pt,y=0.39pt,yscale=-1,xscale=1]
        \draw   (170,2325) .. controls (170,2305.67) and (185.67,2290) .. (205,2290) .. controls (224.33,2290) and (240,2305.67) .. (240,2325) .. controls (240,2344.33) and (224.33,2360) .. (205,2360) .. controls (185.67,2360) and (170,2344.33) .. (170,2325) -- cycle ;
        \draw   (90,2205) .. controls (90,2185.67) and (105.67,2170) .. (125,2170) .. controls (144.33,2170) and (160,2185.67) .. (160,2205) .. controls (160,2224.33) and (144.33,2240) .. (125,2240) .. controls (105.67,2240) and (90,2224.33) .. (90,2205) -- cycle ;
        \draw    (100.32,2235.08) -- (63.28,2289.72) ;
        \draw [shift={(61.6,2292.2)}, rotate = 304.13] [fill={rgb, 255:red, 0; green, 0; blue, 0 }  ][line width=0.08]  [draw opacity=0] (8.93,-4.29) -- (0,0) -- (8.93,4.29) -- cycle    ;
        \draw [shift={(102,2232.6)}, rotate = 124.13] [fill={rgb, 255:red, 0; green, 0; blue, 0 }  ][line width=0.08]  [draw opacity=0] (8.93,-4.29) -- (0,0) -- (8.93,4.29) -- cycle    ;
        \draw    (167,2325) -- (83,2325) ;
        \draw [shift={(80,2325)}, rotate = 360] [fill={rgb, 255:red, 0; green, 0; blue, 0 }  ][line width=0.08]  [draw opacity=0] (8.93,-4.29) -- (0,0) -- (8.93,4.29) -- cycle    ;
        \draw [shift={(170,2325)}, rotate = 180] [fill={rgb, 255:red, 0; green, 0; blue, 0 }  ][line width=0.08]  [draw opacity=0] (8.93,-4.29) -- (0,0) -- (8.93,4.29) -- cycle    ;
        \draw   (10,2325) .. controls (10,2305.67) and (25.67,2290) .. (45,2290) .. controls (64.33,2290) and (80,2305.67) .. (80,2325) .. controls (80,2344.33) and (64.33,2360) .. (45,2360) .. controls (25.67,2360) and (10,2344.33) .. (10,2325) -- cycle ;
        \draw  [dash pattern={on 0.84pt off 2.51pt}]  (149.66,2235.1) -- (188,2293) ;
        \draw [shift={(148,2232.6)}, rotate = 56.49] [fill={rgb, 255:red, 0; green, 0; blue, 0 }  ][line width=0.08]  [draw opacity=0] (8.93,-4.29) -- (0,0) -- (8.93,4.29) -- cycle    ;
        \draw  [dash pattern={on 0.84pt off 2.51pt}]  (83,2310) -- (170,2310) ;
        \draw [shift={(80,2310)}, rotate = 0] [fill={rgb, 255:red, 0; green, 0; blue, 0 }  ][line width=0.08]  [draw opacity=0] (8.93,-4.29) -- (0,0) -- (8.93,4.29) -- cycle    ;
        
        \draw (205,2325) node   [align=left] {SP};
        \draw (125,2205) node   [align=left] {IdP};
        \draw (45,2338) node   [align=left] {User};
        \draw (45,2312) node   [align=left] {IdP};
    \end{tikzpicture}
    \label{fig:iam:ssi}
}
    \caption{The IAM models. Constant lines represent interactions, and dashed lines mean trust.}
    \label{fig:iam}
\end{figure}

Figures~\ref{fig:iam:centralized}, \ref{fig:iam:outsourced} and \ref{fig:iam:usercentric} illustrate these three identity models, providing an overview of the interactions between the user, IdP, and SP. The emergence of specialized IAM services, \textit{i.e.}, third-party IdPs, resulted in the formation of electronic identity oligopolies~\cite{ingram2017facebook}. Long-term users of IdPs are effectively imprisoned by them, as IdPs do not support portability. These companies promote their own rules, which can result in a user being removed from their platforms if those rules are violated. This can be devastating for individuals who have spent years developing trusting relationships with SPs. They will lose their transaction history and become completely unknown if they are banned. This issue is particularly noticeable for IdPs that double as social media platforms, such as Facebook, LinkedIn, and Twitter, where violations of social network rules are often questionable~\cite{eff2015}.

\subsection{Self-Sovereign Identity}

In the early days of the web, the conception of the client-server model shaped the idea that in the digital world, people are users of online systems rather than human beings, \textit{i.e.}, entities that need identification, authentication, and authorization to access and perform tasks online~\cite{searls2012}. This digital model assumes administrative precedence because it was built on the foundation that servers (companies, online businesses) are more important than clients (individuals) and, therefore, dictate the rights of clients~\cite{loffreto2013administrative}. This web fabric holds to this day and is exacerbated by the need for the creation of legislation, such as the European Union's General Data Protection Regulation (GDPR)~\cite{gdpr2016} and the California Consumer Privacy Act (CCPA)~\cite{ccpa2018}, to specify the rights of individuals and their digital data in a society increasingly dependent on digital interactions.

The fundamental premise of SSI is that individuals have sovereignty over their digital selves and thus control over their data. This concept fundamentally distinguishes SSI from previous identity models, which viewed individuals as users. In this new model, sovereign individuals store and manage their data, thereby controlling with whom their private data are shared and to what extent.

Although philosophers such as John Locke and Stuart Mill have written about the sovereignty of individuals in past centuries~\cite{mill1885liberty,sep-locke}, Loffreto~\cite{loffreto2012sovereign} established the first widely accepted~\cite{Path2016,searls2018,kaliya2021,vescent2019comprehensive,young2020domains} link between sovereignty and digital identity~\cite{loffreto2012sovereign}. Thereafter, the meaning of sovereign identity was debated~\cite{searls2013,searls2013a,searls2014,loffreto2013recalibrating}, and technology standards were proposed~\cite{w3cVC,w3cDID}. Significant momentum was obtained, especially in academia~\cite{Muhle,Toth}, after Christopher Allen laid out what he proposed to be the ten principles of SSI~\cite{Path2016}, which are detailed next.

First, individuals must have an \emph{existence} independent of their digital selves, \textit{i.e.}, they cannot exist only virtually. A (self-sovereign) identity works by sharing the desired (digital) aspects of the individual. Second, people must \emph{control} their identities by owning and managing their attributes, which does not prohibit them from making \emph{claims} about other people. Third, people must have \emph{access} to their data and claims by storing them or being readily available if they are outsourced. Fourth, all systems must be \emph{transparent} and the underlying algorithms must be free and open-source, thus allowing detailed examination by anyone. Fifth, identities must \emph{persist} forever, or as long as individuals wish. Sixth and seventh, identities and their claims must be \emph{portable} across different systems and technologies, which requires \emph{interoperability} between standards and implementations. Eighth and ninth, people need to \emph{consent} to the use and sharing of their data, while data disclosure must be \emph{minimized} to the absolute minimum. For instance, to find out if a person can buy an alcoholic beverage, it is unnecessary to share their date of birth. Tenth, at the end of the day, individuals' rights must be \emph{protected}, which means that systems must be designed to avoid censorship and to protect individuals' rights, even at the expense of the system.

In SSI, any assertion about a subject is referred to as a \emph{claim}. A \emph{credential} is a collection of one or more assertions made about a subject by an entity. It could be, for example, a government-issued driver's license that contains a person's date of birth, name, and address. A \emph{Verifiable Credential (VC)} is a credential that includes a revocation list or another method of revocation and contains cryptographic material that ensures the credential's integrity, as well as the issuer's identification and non-repudiation~\cite{w3cVC}. Additionally, a tamper-resistant claim derived from a verifiable credential is referred to as a \emph{verifiable claim} or \emph{Verifiable Presentation (VP)}. Although we use these terms interchangeably throughout this paper, we refer to tamper-proof claims and tamper-proof credentials.

In the same way that entities issue physical credentials to holders in the form of paper or plastic cards in the physical world, entities issue VCs to holders in SSI. However, unlike physical and digital identities, these electronic documents enable individuals to select which attributes (claims) to share, which is impossible with physical or digital credentials. They require the holder to present the identity document in its entirety, revealing all of its attributes. 

Suppose that you are asked to prove that you have reached the age of majority. With a physical document, showing the paper or plastic card will reveal the birthdate and all other attributes to the RP. The same is true for digital identity documents, which are commonly implemented using X.509 attribute certificates~\cite{farrell2010rfc}. With traditional X.509 certificates, the whole certificate has to be shared with the RP to verify the document's integrity. However, in the context of SSI, you would construct a VP stating that: (i) a credential was issued to you by a trusted party; (ii) this credential has your birthdate in it; (iii) your birthdate was more than 18 years ago; and (iv) this credential has not been revoked by the government body. Hence, whoever receives this VP does not learn your name, birthdate, and any other information in the credential, only that you have reached the age of majority.

The recipient of a VP (\textit{i.e.}, the RP) verifies the following: (i) who signed the credential that supports this VP; (ii) whether the VP is constructed correctly (\textit{i.e.}, it contains the required information and is not corrupted or counterfeited); and (iii) whether the credential that supports this VP is valid (\textit{i.e.}, whether the credential was revoked or not). It is important to note that once the issuer of the credential has been verified in step (i), the RP is free to decide whether or not to trust the issuer. Moreover, step (iii) does not require the RP to inquire the IdP in any particular manner. Revocation registries are publicly available, and the verification is done anonymously~\cite{camenisch2001efficient,camenisch2002dynamic}, that is, without disclosing the credential's unique identifier.

While individuals in SSI have the autonomy to issue their own credentials, others are free to distrust them. For example, a bank is unlikely to accept the VP of a self-issued credential that contains a person's name and birthdate. This is true in both the real and virtual worlds. The diagram in Figure~\ref{fig:iam:ssi} depicts a high-level overview of SSI in which the user (\textit{i.e.}, the holder) can interact with the SP using either self-issued or third-party-issued credentials. In either case, the SP is free to decide whether or not to trust the issuer.

Despite the SSI literature's use of the term VP, this concept predates SSI by many years. Prior to SSI, more than a decade of research had been conducted on how to share portions of a credential, as well as predicates over one or more attributes, without losing integrity and authenticity~\cite{chaum1985security,camenisch2001efficient}. Zero-Knowledge Proof (ZKP) is the primary technique underlying VP~\cite{lee2021privacy,Lee2019,Schanzenbach2019}. In short, a ZKP enables a prover to convince a verifier that she is aware of a value without disclosing the value~\cite{schnorr1989efficient}. By combining ZKP and credentials, a credential holder can establish the validity and content of one or more credentials without disclosing the entire credential~\cite{Lee2019}. The same is true for a VC's status. It is possible to demonstrate that a VC has not been revoked without disclosing the credential to the RP and without informing the issuer that a query for a specific credential was made~\cite{camenisch2002dynamic}.

\begin{figure}[H]
    \centering
    \input{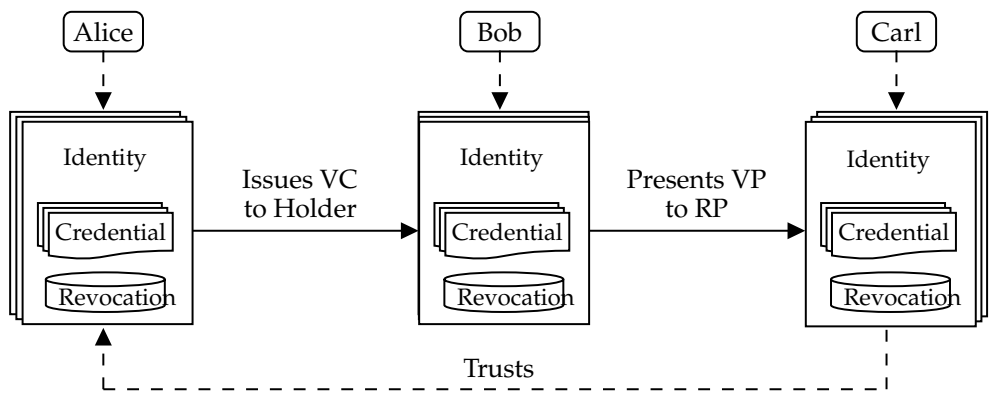}
    \caption{The actors, their electronic identities, and the interactions to issue a VC and present a VP.}
  \label{fig:interaction}
\end{figure}

In Figure~\ref{fig:interaction}, we illustrate the end-to-end process of issuing a VC and emitting a VP in a simplified three-actor model. In this example, three individuals own and control their electronic identities, each of which is appropriate for a particular situation. Each electronic identity is linked to a database of issued and received credentials, as well as a revocation registry for expired or revoked credentials. One of Alice's electronic identities issues a credential to one of Bob's electronic identities, such as a declaration that he is a reputable seller of fine wines. Bob then creates and sends a VP to Carl, proving that he possesses a credential attesting to his good reputation. Carl has trust in the issuer of the credential from which that VP was derived, Alice, an internationally renowned winemaker. Carl then begins negotiating with Bob. It should be noted that, in reality, the majority of people will not host revocation registries because they do not issue credentials, which is also the case for physical and digital identification documents.

This simplified example demonstrates the trust mechanics of SSI. However, it lacks the depth and complexities of real-life scenarios. For instance, a user may create a VP using two credentials, one of which is deemed trustworthy while the other is not. Deriving trust in non-trivial scenarios is one of the open challenges in SSI.

After discussing electronic identities and the evolution of IAMs, we introduced the reader to SSI. Following that, in Section~\ref{sec:related}, we present other SSI surveys and their shortcomings, followed by the method used in this systematic review and systematic mapping in Section~\ref{sec:method}.

\section{Related Work}\label{sec:related}
Blockchain technology pioneered the concept of distributed ledgers, in which peer consensus defines the immutable ledger's state rather than a central entity asserting authoritarian control~\cite{swan2015blockchain}. These concepts facilitate the implementation of SSI by providing a trusted online repository for electronic identities, credentials, and revocation registries~\cite{Kuperberg2019}. While blockchain technology can help the development of SSI solutions, it is not required~\cite{Alboaie2017a,alpar2017irma,van2019self,Linklater2018b}. Despite this, the majority of existing reviews claim that SSI cannot be implemented without blockchain~\cite{Kuperberg2019,Liu2020b,zhu2018identity,Lim2018,kaneriya2020comparative,gilani2020survey,dib2020decentralized,Muhle}. In terms of research method, one of the existing surveys conducted a systematic mapping of the literature~\cite{vcuvcko2021decentralized}, two devised taxonomies~\cite{ghaffari2021identity,schmidt2021clear}, one carried a meta-synthesis~\cite{mulaji2021practicality}, and seven did not detail any method for selecting and analyzing primary sources~\cite{Kuperberg2019, zhu2018identity,Lim2018,kaneriya2020comparative,gilani2020survey,dib2020decentralized,Muhle}. Next, we present existing secondary research in SSI.

Kuperberg~\cite{Kuperberg2019} conducted a survey in which forty-three blockchain-based SSI market offerings were evaluated against seventy-five criteria, including compliance with applicable legislation, market availability, and cost. He stipulated that no reviewed application meets all criteria, and no SSI solution possesses the following characteristics: (i) the maturity of traditional IAM offerings; (ii) a production-level integration standard (such as OAuth 2.0~\cite{hardt2012oauth} or SAML~\cite{hughes2005security}); and (iii) OS-level integration.

Although Liu \textit{et al.}~\cite{Liu2020b} presented their search string, they do not provide any information about their review method. Thirty-six research efforts and patents introducing SSI applications are reviewed in total. They examined these works from the standpoints of authentication, privacy, and trust. They argued that, despite blockchain-related innovations, there are still issues and implications remaining, namely: (i) users may lose their blockchain-based identities (wallets) and need to (ii) change their identities, which is trivial in traditional IAM but might be challenging in distributed ledgers; and (iii) the cost of integrating existing systems into the new paradigm.

Zhu and Badr~\cite{zhu2018identity} conducted a review of works that use distributed ledgers to implement SSI in the context of IoT devices. They expanded on Liu \textit{et al.}'s~\cite{Liu2020b} focus on authentication, privacy, and trust, adding a fourth dimension: performance. They alleged that the trustless environments in which IoT devices operate necessitate SSI solutions. Nonetheless, blockchain technology should be thoroughly investigated, as storing and maintaining public blockchains in IoT devices is prohibitively resource-intensive. As a result, forming small groups of private blockchains may be an option. According to the literature, one possible solution is for IoT devices to inherit the peer-to-peer trust established between their owner entities (humans, businesses, and governments)~\cite{zhu2018fog}.

Despite the comparison of the underlying infrastructure of blockchain-based SSI offerings, three surveys that do not specify a search method produced similar results~\cite{Lim2018,kaneriya2020comparative,gilani2020survey}. They all mentioned the blockchain framework that the surveyed papers use, as well as the type of blockchain network (private, permissioned, permissionless, or other). Lim \textit{et al.}~\cite{Lim2018} conducted a review of 15 for-profit and non-profit company-developed, government-related, and open-source applications, concluding that SSI is the optimal solution for user-centric, secure, and cost-effective IAM. Kaneriya and Patel~\cite{kaneriya2020comparative} conducted a review of six SSI systems, identifying future enhancements that each system, according to the authors, should prioritize. Finally, Gilani \textit{et al.}~\cite{gilani2020survey} reviewed eight SSI offerings, noting which support selective disclosure of personal information, how cryptographic keys are managed, and blockchain-specific details such as whether credentials are stored on or off the ledger, as well as the use of smart contracts. Smart contracts is a software that executes automatically and transparently on the ledger, allowing anyone to verify them~\cite{zheng2020overview}.

The authors of~\cite{dib2020decentralized} described ten SSI systems that utilize blockchain technology but did not specify how they were chosen. They did, however, conduct an analysis of these works in terms of their adherence to the SSI's ten principles, detailing which principle each reviewed paper satisfies. 

In contrast to previous surveys, Mühle \textit{et al.}~\cite{Muhle} examined what they refer to as the ``four basic components of SSI'': identification; authentication; verifiable claims; and attribute storage. They discussed how various research studies and market offerings attempt to address each of the four components.

Čučko \textit{et al.}~\cite{vcuvcko2021decentralized} presented a systematic map of decentralized identity. They mapped one hundred and twenty papers in total, but only eighty were determined to be SSI-related. While they established a category for conceptual contributions, it was filled up with surveys and research articles highlighting SSI's challenges and opportunities. Alternatively, we consider conceptual contributions that refute or include new philosophical perspectives on what SSI is. Their map encompasses information technology fields and the various domains to which SSI is applied, whereas our maps depict the relationship between authors and publications.

Taxonomies for SSI are introduced by both~\cite{ghaffari2021identity} and~\cite{schmidt2021clear}. The former proposes a four-tiered taxonomy encompassing registration, authentication, data management, and verifiable claims. They were used to categorize twenty-one blockchain-based solutions. The latter's taxonomy includes the facets member, interaction, ambition, and technology stack, which are used to classify one hundred forty-seven results from a gray literature review of the SSI ecosystem culled from DuckDuckGo, Github, Reddit, and ArXiv. Both taxonomies fall short of incorporating philosophical debates about the meaning of SSI.

Finally, the authors of~\cite{mulaji2021practicality}  created a meta-synthesis of SSI based on blockchain technology. Meta-synthesis is a qualitative method for aggregating knowledge derived from quantitative, qualitative, empirical, conceptual, and review studies~\cite{finfgeld2018guide}. They evaluated sixty-nine works from an enterprise adoption perspective, summarizing the state of the art's technological and business challenges.

Secondary research has already revealed an increasing number of studies in this field. However, a rigorous systematic review of SSI studies is lacking. Earlier studies have examined both the practical and technical aspects of SSI systems. However, they do not evaluate conceptual debates about SSI or works that present and attempt to resolve particular pragmatic issues. On the other hand, we are interested in discovering and examining research materials that extend or refute Allen's ten principles of self-sovereign identity~\cite{Path2016} or present and resolve practical problems in the SSI ecosystem. Table~\ref{tbl:surveys} summarizes the major differences between previous surveys and ours.

\captionsetup[table]{singlelinecheck=off}

\begin{table}[H]
    \caption{Comparison with other secondary studies in the literature.}
    \label{tbl:surveys}
    \begin{adjustwidth}{-\extralength}{0cm}
        \begin{tabular}{l c c c c c c c}
            \toprule
            & \makecell{\bf Systematic\\ \bf Review} & \makecell{\bf Systematic\\ \bf Mapping} & \bf Taxonomy & \makecell{\bf Include\\ \bf Patents} & \makecell{\bf Other than\\ \bf Blockchain} & \makecell{\bf Conceptual \\ \bf or Pragmatic} & \makecell{\bf Covered \\ \bf Works} \\ \midrule
            Kuperberg~\cite{Kuperberg2019} & No & No & No & No & No & Pragmatic & 43 \\
            Liu \textit{et al.}~\cite{Liu2020b} & No\tablefootnote{Presented their search string but do not provide any information about their review method.} & No & No & Yes & No & Pragmatic & 50 \\
            Zhu and Badr~\cite{zhu2018identity} & No & No & No & No & No & Pragmatic & 15 \\
            Lim \textit{et al.}~\cite{Lim2018} & No & No & No & No & No & Pragmatic & 15 \\
            Kaneriya and Patel~\cite{kaneriya2020comparative} & No & No & No & No & No & Pragmatic & 6 \\
            Gilani \textit{et al.}~\cite{gilani2020survey} & No & No & No & No & No & Pragmatic & 8 \\
            Dib and Toumi~\cite{dib2020decentralized} & No & No & No & No & No & Pragmatic & 10 \\
            Mühle \textit{et al.}~\cite{Muhle} & No & No & No & No & No & Pragmatic & 9 \\
            Čučko \textit{et al.}~\cite{vcuvcko2021decentralized} & No & Yes & No & No & Yes & Pragmatic & 80 \\ 
            Ghaffari \textit{et al.}~\cite{ghaffari2021identity} & No & No & Yes & No & No & Pragmatic & 21 \\ 
            Schmidt \textit{et al.}~\cite{schmidt2021clear} & No & No & Yes & No & No & Pragmatic & 147 \\  
            Mulaji and Roodt~\cite{mulaji2021practicality} & No & No & No & Yes & No & Pragmatic & 69 \\  \midrule
            \textbf{This work} & \textbf{Yes} & \textbf{Yes} & \textbf{Yes} & \textbf{Yes} & \textbf{Yes} & \textbf{Both} & \textbf{82} \\
            \bottomrule
        \end{tabular}
    \end{adjustwidth}
\end{table}

\section{Method}\label{sec:method}
Secondary studies are necessary to keep track of advancements and developments as primary research efforts on a given topic evolve. Two types of secondary studies have gained popularity in recent years in computer science~\cite{Petersen2015}: systematic mapping~\cite{Petersen2008} and systematic literature review~\cite{KitchenhamB.andCharters2007}. Despite the fact that both are systematic and thus employ rigorous methods for identifying and interpreting relevant research, the former is intended to provide a broad overview and identify research trends, whereas the latter is intended to aggregate evidence in order to summarize and answer more specific Research Questions (RQs). In this study, we conducted a systematic review of the literature and a systematic mapping.

\subsection{Planning}

We followed Petersen~\textit{et al.}'s method~\cite{Petersen2015}, which provides detailed guidelines based on a systematic review of mapping studies. These guidelines require the following: (i) the definition of objectives and RQs; (ii) a strategy for identifying relevant studies; (iii) objective inclusion and exclusion criteria to ensure that only relevant material is reviewed; (iv) an extraction process for objectively obtaining evidence from papers relevant to the RQs; (v) a classification method; and (vi) a discussion of potential threats to the study's validity. Our research protocol, which is detailed in the following sections, complies with the aforementioned stipulations.

\subsubsection{Research Questions}\label{sec:RQs}

The objective of this systematic study is fourfold: (i) to examine practical challenges associated with SSI and potential solutions; (ii) to investigate mathematical formalism and cryptographic tools (primitives) used to solve these problems; (iii) to investigate conceptual advancements made to the informal definition of SSI~\cite{Path2016}; and (iv) to map SSI publications and authors. These goals result in the following RQs:

\begin{enumerate}[label=\textbullet\text{  }RQ-\arabic*:,labelindent=\parindent,leftmargin=*,labelsep*=0.1em, ref=RQ-\arabic*]
    \item What practical problems have been introduced and solved? \label{rq-1}
    \item What properties, formal definitions and cryptographic tools have been used? \label{rq-2}
    \item What conceptual ideas have been introduced or refuted? \label{rq-3}
    \item When, where, and by whom were SSI studies published? \label{rq-4}
\end{enumerate}

\subsubsection{Search Strategy}\label{sec:search-strategy}

Our investigation began by specifying a search string that was pertinent to the RQs previously mentioned. Rather than creating a potentially restrictive search query using PICOC~\cite{KitchenhamB.andCharters2007} or another method of query framing, we searched for ``self-sovereign identity'' and variants in the title, author keywords, and abstract. Our search string is broad by design in order to encompass as many relevant articles, patents, and research materials as possible. Additionally, we placed no restrictions on the publication year, page count, conference, or journal. The following is the entirety of our query string.

\begin{figure}[H]
    \resizebox{\linewidth}{!}{
    \fbox{self-sovereign identity \textbf{OR} self sovereign identity \textbf{OR} self-sovereignty \textbf{OR} self sovereignty}
    }
\end{figure}

\subsubsection{Study Selection}\label{sec:study-selection}

Our study selection process is divided into three stages. The first phase eliminates duplicate results and articles that have been republished in extended formats. Mendeley~\cite{mendeley} was used to evaluate the results and eliminate duplicates.

After a preliminary screening of the search results, it was determined that several papers do not belong in the field of computer science or are not relevant to our review. We then narrowed our search by developing two inclusion criteria and one exclusion criterion. These criteria are detailed in Table~\ref{tbl:ICEC}. In short, the exclusion criterion eliminates research that is not computer science-related, whereas the inclusion criterion prioritizes papers that contribute to SSI in response to our RQs. Articles had to meet at least one inclusion criteria.

\captionsetup[table]{singlelinecheck=on}
\begin{table}[H]
    \centering
    \caption{Inclusion and exclusion criteria.}
    \label{tbl:ICEC}
    \begin{tabular}{l l}
        \toprule
        \multicolumn{2}{l}{\textbf{Inclusion Criteria}} \\ \midrule
        IC-1 & The paper includes a novel conceptual contribution to SSI. \label{IC-1} \\
        IC-2 & The research work makes practical progress towards SSI. \label{IC-2} \\ 
        \midrule
        \multicolumn{2}{l}{\textbf{Exclusion Criterion}} \label{EC-1} \\ \midrule
        EC-1 & The research work is not in the area of computer science. \\
        \bottomrule
    \end{tabular}
\end{table}

We are not reviewing and mapping standalone SSI solutions, despite the fact that they may incorporate practical progress (such as Sovrin~\cite{sovrin} and uPort~\cite{uport}). Multiple surveys have been conducted on these works~\cite{Lim2018,Kuperberg2019,Liu2020b,zhu2018identity}. 
As a result, when it comes to practical progress, we prioritize works that raise specific pragmatic concerns about any aspect of the SSI ecosystem and propose solutions. Consider, for example, a piece that discusses the difficulty of recovering SSI keys that have been lost and offers a new solution to the problem. This work would comply with \hyperref[IC-2]{IC-2}. Assume, however, that a research paper is published describing an implementation of SSI for IoT. While this work may make a significant contribution to the IoT literature, it does not satisfy \hyperref[IC-2]{IC-2} if it does not present a problem concerning SSI in general and a solution to that problem.

\hyperref[EC-1]{EC-1} is applied to the title, author keywords, and abstract in the second stage of our study selection process, effectively eliminating articles that are not related to computer science. The third phase involves obtaining and reading the remaining studies in their entirety, ensuring that they comply with \hyperref[IC-1]{IC-1}, \hyperref[IC-2]{IC-2}, or both. Then, articles that violate \hyperref[IC-1]{IC-1} or \hyperref[IC-2]{IC-2} are removed as well.

\subsubsection{Data Extraction}\label{sec:data-extraction}

To extract data from primary studies, we adapted Petersen's template~\cite{Petersen2015}. It is composed of three components: (i) a data item; (ii) a description; and (iii) the RQ to which the data item corresponds, as illustrated in Table~\ref{tbl:extraction_form}. Except for the Study ID, which was generated manually, the \emph{General} items were obtained from articles or their online metadata. Following the reading of a pilot set of articles, two \emph{Conceptual} and two \emph{Practical} data items were created to gather evidence and address the RQs.

\begin{table}[H]
    \centering
    \caption{Data extraction form adapted from~\cite{Petersen2015}.}
    \label{tbl:extraction_form}
    \begin{tabular}{lll}
        \toprule
        \textbf{Data Item} & \textbf{Description} & \textbf{RQ} \\ \midrule
        \textit{General}\label{general} & & \\
        Study ID & Unique integer identifier per article &  \\
        Article Title & Name of the article &  \\
        Year & Year of publication & \ref{rq-4} \\
        Article Authors & Name of the authors & \ref{rq-4} \\
        Venue & Publication venue & \ref{rq-4} \\
        \midrule
        \textit{Conceptual} & & \\
        Add Concept\label{add-concept} & What concept/idea is introduced & \ref{rq-3} \\
        Refute Concept\label{refute-concept} & What concept/idea is refuted & \ref{rq-3} \\
        \midrule
        \textit{Formalism} & & \\
        Formal Model\label{formal-model} & How is SSI 
        formally specified & \ref{rq-2} \\
        \midrule
        \textit{Practical} & & \\
        Novel Problem\label{novel-problem} & What practical problem is presented & \ref{rq-1} \\
        Proposed Solution\label{proposed-solution} & How is the practical problem solved & \ref{rq-1} \\
        \bottomrule
    \end{tabular}
\end{table}

\subsubsection{Taxonomy}\label{sec:classification}

To develop a taxonomy to categorize SSI research, we used the three-step keywording method~\cite{Petersen2008}: (i) the researcher reads the abstracts (and, if the abstract is of low quality, the introduction and conclusion as well), extracting keywords and concepts that indicate the article's contribution and the context of the research; (ii) the set of keywords is combined to create a high-level understanding of the research contribution; and (iii) the final set of keywords is clustered to create categories. The last step is the result of the process of making, updating, and merging categories, as well as classifying articles into the new categories that were made.

\subsection{Execution}

\subsubsection{Search Execution}

On February 15, 2022, this search string was entered into the ACM Digital Library~\cite{acm}, IEEE Xplore~\cite{ieee}, ScienceDirect~\cite{sd}, and Springer Link~\cite{springer} databases, which host popular computer science conferences and journals. To supplement our database search, we performed additional searches on Scopus Preview~\cite{scopus}, Web of Science~\cite{wos}, and Google Scholar~\cite{scholar}. Additionally, we queried Google Patents~\cite{googlePatents} on the same day and applied the search string to the title and abstract of patents, yielding seventeen results. Table~\ref{tbl:execution} displays the number of search results returned by the queries.
%

\begin{table}[H]
    \centering
    \caption{Number of studies.}
    \label{tbl:execution}
    \begin{tabular}{ l r }
        \toprule
        \textbf{Tool} & \textbf{Total} \\ \midrule
        ACM Digital Library & 16 \\ 
        IEEE Xplore Digital Library & 99 \\ 
        ScienceDirect & 17 \\
        Springer Link & 40 \\ 
        Scopus & 235 \\ 
        Web of Science & 131 \\ 
        Google Scholar & 180 \\ \midrule
        \textbf{Database Search} & \textbf{718} \\ \midrule
        Google Patents & 17 \\ \midrule
        \textbf{Patent Search} & \textbf{17} \\
        \bottomrule
    \end{tabular}
\end{table}

\subsubsection{Study Selection and Data Extraction}

\begin{figure}[H]
    \centering
    \input{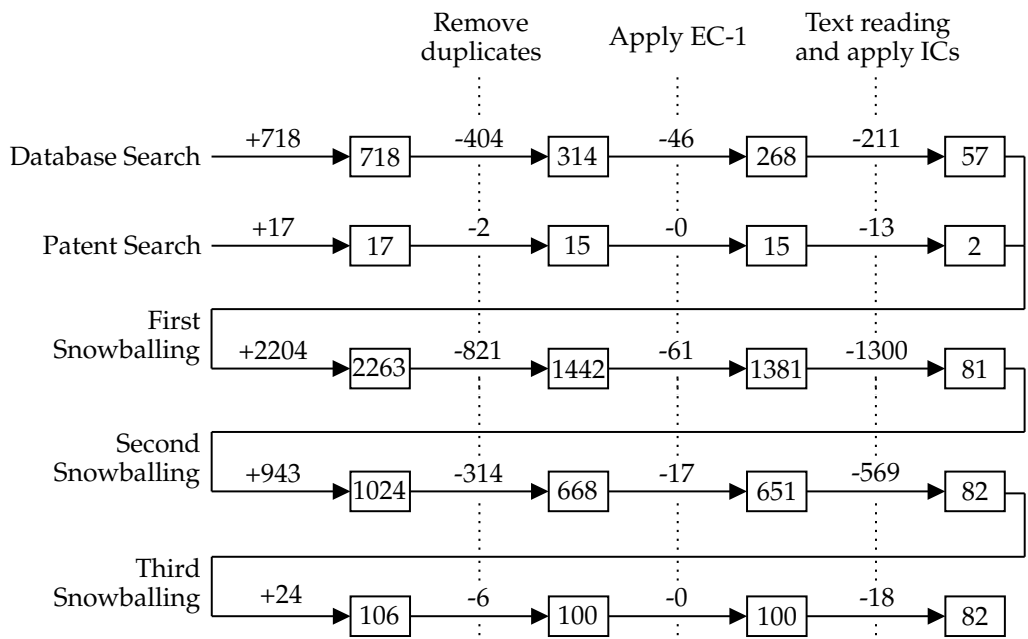}
    \caption{Number of articles in each stage of our study selection.}
    \label{fig:search_selection}
\end{figure}

Our three-phase study selection process was executed five times, as presented in  Figure~\ref{fig:search_selection}. We applied the first execution to the outputs of the database search and the second to the patent search results. The combined output was a set of fifty-nine works which formed the input set for both forward and backward snowballing~\cite{Wohlin2014}. In short, backward snowballing consists of reviewing all references in a document, while forward snowballing finds other works that reference it. The snowballing was repeated until no new work was found that satisfied our selection process, which required three runs. The remaining eighty-two works constitute our result set. We should point out that two researchers independently assessed each paper at every stage of the selection process, and a conflict resolution meeting was organized. We point the interested reader elsewhere~\cite{spreadsheet} for the complete list of papers, our evaluation regarding their inclusion or exclusion for all five runs of the study selection process, and the data extracted with our data collection form.

\subsubsection{Threats to Validity}\label{sec:threats}

The following validity threats are critical and must be highlighted~\cite{Petersen2015}: (i) descriptive; (ii) study identification; and (iii) data extraction and classification.

To mitigate the risk of collecting observations inaccurately from research papers, \textit{i.e.}, the descriptive validity threat, we developed and used the data collection form described above to collect relevant evidence. The first author used the data collection form, and the second author evaluated the results.

Following that, to minimize the possibility of overlooking relevant work, \textit{i.e.}, the study identification validity threat, we did not restrict our database search by publication year or venue. Backward and forward snowballing was also used to supplement the database search.

Concerning the last threat to validity, namely data extraction and classification threat, it should be noted that researcher bias and human error cannot be completely eliminated because these processes involve human judgment. To avoid this, the second author examined at the first author's data extraction and classification.

Furthermore, it is worth stressing that identity management has been extensively studied for decades. Thus, despite the fact that numerous research efforts were conducted before the term ``self-sovereign identity'' was coined, a large number of research efforts can arguably contribute to the many facets of SSI. Ultimately, deciding which work makes a significant contribution to SSI is entirely dependent on the researcher's interpretation. To avoid this interpretation bias, we reviewed and mapped works that explicitly mention the term self-sovereign identity or any synonym from our search string.

\subsubsection{Findings}

The next five sections present our findings. First, the proposed taxonomy is introduced. Then, the following four sections answer our RQs respectively.


\section{Taxonomy of Self-Sovereign Identity}\label{sec:taxonomy}
\begin{figure}[h]
    \centering
        \resizebox{\linewidth}{!}{
        \input{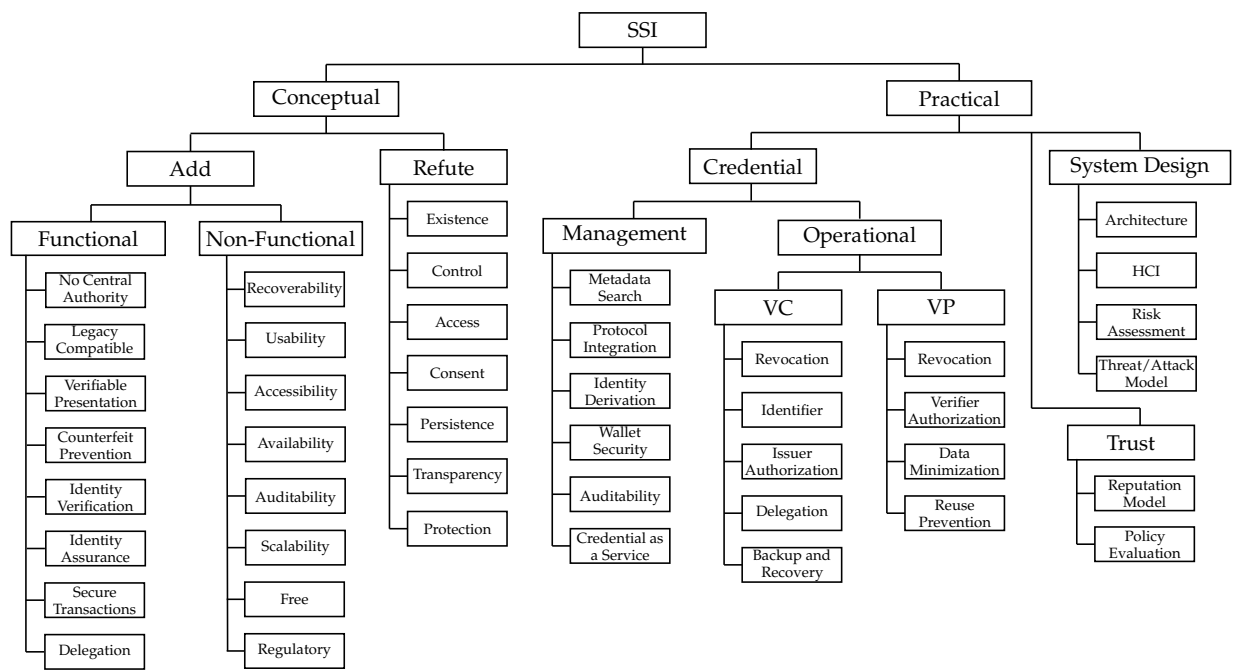}
        }
        \caption{Taxonomy of SSI.}
        \label{fig:tax}
\end{figure}

We used the keywording method~\cite{Petersen2008} to identify distinguishing characteristics of the reviewed work. These characteristics were combined into a proposed taxonomy with two facets: \emph{conceptual} and \emph{practical}, as illustrated in Figure~\ref{fig:tax}. These two facets are further subdivided into additional facets, forming a tree-like hierarchy. Concepts, sometimes referred to as terms, are the leaves of this hierarchical tree.

The \emph{conceptual} facet categorizes the research efforts that, during our data extraction process, filled in the data items \hyperref[add-concept]{\emph{Add Concept}} or \hyperref[refute-concept]{\emph{Refute Concept}} and thus help answer~\ref{rq-3}. The new concepts are divided into two facets: \emph{functional}, which refers to the well-defined functionalities of SSI systems; and \emph{non-functional}, which refers to more generic behaviors.

The \emph{practical} facet is used to classify publications that make pragmatic contributions, \textit{i.e.}, those that contribute to the data items \hyperref[novel-problem]{\emph{Novel Problem}} and \hyperref[proposed-solution]{\emph{Proposed Solutions}}, and thus related to~\ref{rq-1}. It is divided into three facets that are used to analyze work that presents challenges and proposes solutions in the following areas: (i) \emph{management} and \emph{operational} aspects of \emph{credentials}; (ii) \emph{system design}; and (iii) \emph{trust}. The \emph{operational} facet is further subdivided into the \emph{VC} and \emph{VP} facets.

The number of existing concepts under the facets of our proposed taxonomy, \textit{i.e.}, the leaves, is likely to grow in the future. New research, for example, may introduce new pragmatic challenges. Future work can build on our proposed taxonomy and include new initiatives.

We present and discuss the state-of-the-art of SSI in the following sections through the lens of the proposed taxonomy. These sections are arranged in accordance with the taxonomy's facets and concepts. We discuss them and the works in terms of their most defining facet, namely the objective or problem they are attempting to solve because the majority of surveyed works are classified under multiple facets due to exhibiting a variety of characteristics. We begin with the \emph{practical} facet.

\section{RQ-1: What practical problems have been introduced and solved?}\label{sec:RQ1}

Our taxonomy enabled us to classify surveyed materials and generate visualizations to help answer our research questions. The data items in our data extraction form pertaining to our first research question are organized in Table~\ref{tbl:R1} according to the facets and terms of our taxonomy under the \emph{practical} facet, which were fulfilled by sixty-nine of the eighty-two reviewed materials. 

\captionsetup[table]{singlelinecheck=off}

\begin{table} \centering
    \begin{adjustwidth}{-\extralength}{0cm}
    \caption{Publications that introduced and solved novel problems in the SSI ecosystem.}
    \label{tbl:R1}
    \resizebox*{\linewidth}{!}{
        \begin{tabular}{rcccccclccccclcccclcccclcc}
            \toprule
            & \multicolumn{17}{c}{Credential} && \multicolumn{4}{c}{\multirow{3}{*}[-0.55em]{\centering System Design}} && \multicolumn{2}{c}{\multirow{3}{*}[-0.55em]{Trust}} \\ \cmidrule{2-18}
            & \multicolumn{6}{c}{\multirow{2}{*}[-0.25em]{Management}} && \multicolumn{10}{c}{Operational} &&&&&& \\ \cmidrule{9-18}
            &&&&&&&& \multicolumn{5}{c}{VC} && \multicolumn{4}{c}{VP} &&&&&& \\ \cmidrule{2-7} \cmidrule{9-13} \cmidrule{15-18} \cmidrule{20-23} \cmidrule{25-26}
            Works & \rot{Metadata Search} & \rot{Protocol Integration} & \rot{Identity Derivation} & \rot{Wallet Security} & \rot{Auditability} & \rot{Credential as a Service} &  & \rot{Revocation} & \rot{Decentralized Identifiers} & \rot{Issuer Authorization} & \rot{Delegation} & \rot{Backup and Recovery} &  & \rot{Revocation} & \rot{Verifier Authorization} & \rot{Data Minimization} & \rot{Reuse Prevention} &  & \rot{SSI Design/Architecture} & \rot{HCI} & \rot{Risk Assessment} & \rot{Threat/Attack Model} &  & \rot{Reputation Model} & \rot{Trust Policy Evaluation} \\ \midrule
\cite{Lux2019,schardong2021matching} & \OK &  &  &  &  &  &  &  &  &  &  &  &  &  &  &  &  &  &  &  &  &  &  &  &  \\  \cdashline{2-7} \cdashline{9-13} \cdashline{15-18} \cdashline{20-23} \cdashline{25-26} \addlinespace[1pt]
\cite{anaigoundanpudur2021cryptographic} &  &  &  &  &  &  &  &  &  &  &  &  &  &  & \OK &  &  &  &  &  &  & \OK &  &  &  \\  \cdashline{2-7} \cdashline{9-13} \cdashline{15-18} \cdashline{20-23} \cdashline{25-26} \addlinespace[1pt]
\cite{Schanzenbach2019,kang2021decentralized} &  &  &  &  &  &  &  &  &  &  &  &  &  &  &  &  & \OK &  &  &  &  &  &  &  &  \\  \cdashline{2-7} \cdashline{9-13} \cdashline{15-18} \cdashline{20-23} \cdashline{25-26} \addlinespace[1pt]
\cite{Bathen2019,mishra2021pseudo} &  &  & \OK &  &  &  &  &  &  &  &  &  &  &  &  &  &  &  &  &  &  &  &  &  &  \\  \cdashline{2-7} \cdashline{9-13} \cdashline{15-18} \cdashline{20-23} \cdashline{25-26} \addlinespace[1pt]
\cite{abraham2021ssi} &  &  & \OK &  &  &  &  &  &  &  &  &  &  &  &  &  &  &  &  &  &  & \OK &  &  &  \\  \cdashline{2-7} \cdashline{9-13} \cdashline{15-18} \cdashline{20-23} \cdashline{25-26} \addlinespace[1pt]
\cite{Abraham2020} &  &  & \OK &  &  &  &  & \OK &  &  &  &  &  &  &  & \OK &  &  &  &  &  & \OK &  &  &  \\  \cdashline{2-7} \cdashline{9-13} \cdashline{15-18} \cdashline{20-23} \cdashline{25-26} \addlinespace[1pt]
\cite{Gruner2019,gruner2021atib,martinez2021applying} &  & \OK & \OK &  &  &  &  &  &  &  &  &  &  &  &  &  &  &  &  &  &  &  &  &  & \OK \\  \cdashline{2-7} \cdashline{9-13} \cdashline{15-18} \cdashline{20-23} \cdashline{25-26} \addlinespace[1pt]
\cite{Lux2020} &  & \OK & \OK &  &  &  &  &  &  &  &  &  &  &  &  &  &  &  &  &  &  &  &  &  &  \\  \cdashline{2-7} \cdashline{9-13} \cdashline{15-18} \cdashline{20-23} \cdashline{25-26} \addlinespace[1pt]
\cite{lagutin2019enabling,yildiz2021connecting} &  & \OK &  &  &  &  &  &  &  &  &  &  &  &  &  &  &  &  &  &  &  &  &  &  &  \\  \cdashline{2-7} \cdashline{9-13} \cdashline{15-18} \cdashline{20-23} \cdashline{25-26} \addlinespace[1pt]
\cite{zhong2021jointcloud} &  & \OK &  &  &  &  &  &  &  &  &  &  &  &  &  &  &  &  &  &  &  &  &  & \OK &  \\  \cdashline{2-7} \cdashline{9-13} \cdashline{15-18} \cdashline{20-23} \cdashline{25-26} \addlinespace[1pt]
\cite{Hong2020a} &  & \OK &  &  &  &  &  &  &  &  &  &  &  &  &  &  &  &  &  &  &  & \OK &  &  &  \\  \cdashline{2-7} \cdashline{9-13} \cdashline{15-18} \cdashline{20-23} \cdashline{25-26} \addlinespace[1pt]
\cite{Schanzenbach2018a,Xu2020a} &  & \OK &  &  &  &  &  &  &  &  &  &  &  & \OK &  &  &  &  &  &  &  & \OK &  &  &  \\  \cdashline{2-7} \cdashline{9-13} \cdashline{15-18} \cdashline{20-23} \cdashline{25-26} \addlinespace[1pt]
\cite{schanzenbach2020towards} &  & \OK &  &  &  &  &  &  &  & \OK &  &  &  &  &  &  &  &  &  &  &  & \OK &  &  &  \\  \cdashline{2-7} \cdashline{9-13} \cdashline{15-18} \cdashline{20-23} \cdashline{25-26} \addlinespace[1pt]
\cite{lauinger2021poa} &  &  &  &  &  &  &  &  &  & \OK &  &  &  &  &  &  &  &  &  &  &  & \OK &  &  & \OK \\  \cdashline{2-7} \cdashline{9-13} \cdashline{15-18} \cdashline{20-23} \cdashline{25-26} \addlinespace[1pt]
\cite{muktacredential} &  &  &  &  &  &  &  &  &  & \OK &  &  &  &  &  &  &  &  &  &  &  &  &  &  &  \\  \cdashline{2-7} \cdashline{9-13} \cdashline{15-18} \cdashline{20-23} \cdashline{25-26} \addlinespace[1pt]
\cite{Stokkink2018c} &  &  &  &  & \OK &  &  &  &  &  &  &  &  & \OK &  & \OK &  &  & \OK &  &  &  &  & \OK &  \\  \cdashline{2-7} \cdashline{9-13} \cdashline{15-18} \cdashline{20-23} \cdashline{25-26} \addlinespace[1pt]
\cite{lax2021lightweight} &  &  &  &  &  &  &  & \OK &  &  &  &  &  & \OK &  &  &  &  &  &  &  & \OK &  &  &  \\  \cdashline{2-7} \cdashline{9-13} \cdashline{15-18} \cdashline{20-23} \cdashline{25-26} \addlinespace[1pt]
\cite{w3cVC} &  &  &  &  &  &  &  & \OK &  &  &  &  &  &  &  & \OK &  &  & \OK &  &  &  &  &  &  \\  \cdashline{2-7} \cdashline{9-13} \cdashline{15-18} \cdashline{20-23} \cdashline{25-26} \addlinespace[1pt]
\cite{abraham2020revocable} &  &  &  &  &  &  &  & \OK &  &  &  &  &  &  &  & \OK &  &  &  &  &  &  &  &  &  \\  \cdashline{2-7} \cdashline{9-13} \cdashline{15-18} \cdashline{20-23} \cdashline{25-26} \addlinespace[1pt]
\cite{cho2021verifiable,chotkan2021industry} &  &  &  &  &  &  &  & \OK &  &  &  &  &  &  &  &  &  &  &  &  &  & \OK &  &  &  \\  \cdashline{2-7} \cdashline{9-13} \cdashline{15-18} \cdashline{20-23} \cdashline{25-26} \addlinespace[1pt]
\cite{Lee2019,yang2020zero,lee2021privacy,bobolz2021issuer} &  &  &  &  &  &  &  &  &  &  &  &  &  &  &  & \OK &  &  &  &  &  & \OK &  &  &  \\  \cdashline{2-7} \cdashline{9-13} \cdashline{15-18} \cdashline{20-23} \cdashline{25-26} \addlinespace[1pt]
\cite{jaroucheh2021secretation} &  &  &  &  &  & \OK &  &  &  &  &  &  &  &  &  &  &  &  &  &  &  &  &  &  &  \\  \cdashline{2-7} \cdashline{9-13} \cdashline{15-18} \cdashline{20-23} \cdashline{25-26} \addlinespace[1pt]
\cite{samir2021dt,siddiqui2021credentials} &  &  &  &  &  & \OK &  &  &  &  &  &  &  &  &  &  &  &  &  &  &  & \OK &  &  &  \\  \cdashline{2-7} \cdashline{9-13} \cdashline{15-18} \cdashline{20-23} \cdashline{25-26} \addlinespace[1pt]
\cite{Gruner2019a, bhattacharya2020enhancing,abramson2021evaluating} &  &  &  &  &  &  &  &  &  &  &  &  &  &  &  &  &  &  &  &  &  &  &  & \OK &  \\  \cdashline{2-7} \cdashline{9-13} \cdashline{15-18} \cdashline{20-23} \cdashline{25-26} \addlinespace[1pt]
\cite{Gruner2018} &  &  &  &  &  &  &  &  &  &  &  &  &  &  &  &  &  &  &  &  &  & \OK &  & \OK &  \\  \cdashline{2-7} \cdashline{9-13} \cdashline{15-18} \cdashline{20-23} \cdashline{25-26} \addlinespace[1pt]
\cite{Barclay2020a, Liu2020, Ferdous2019} &  &  &  &  &  &  &  &  &  &  &  &  &  &  &  &  &  &  & \OK &  &  &  &  &  &  \\  \cdashline{2-7} \cdashline{9-13} \cdashline{15-18} \cdashline{20-23} \cdashline{25-26} \addlinespace[1pt]
\cite{WO2021125586A1} &  &  &  & \OK &  &  &  &  &  &  &  &  &  &  &  &  &  &  &  &  &  &  &  &  &  \\  \cdashline{2-7} \cdashline{9-13} \cdashline{15-18} \cdashline{20-23} \cdashline{25-26} \addlinespace[1pt]
\cite{Liu} &  &  &  & \OK &  &  &  &  &  &  & \OK &  &  &  &  &  &  &  & \OK &  &  &  &  &  &  \\  \cdashline{2-7} \cdashline{9-13} \cdashline{15-18} \cdashline{20-23} \cdashline{25-26} \addlinespace[1pt]
\cite{lim2021subject} &  &  &  &  &  &  &  &  &  &  & \OK &  &  &  &  &  &  &  &  &  &  &  &  &  &  \\  \cdashline{2-7} \cdashline{9-13} \cdashline{15-18} \cdashline{20-23} \cdashline{25-26} \addlinespace[1pt]
\cite{lemieux2021addressing} &  &  &  &  & \OK &  &  &  &  &  &  &  &  &  &  &  &  &  &  &  &  &  &  &  &  \\  \cdashline{2-7} \cdashline{9-13} \cdashline{15-18} \cdashline{20-23} \cdashline{25-26} \addlinespace[1pt]
\cite{Jakubeit2020} &  &  &  &  & \OK &  &  &  &  &  &  & \OK &  &  &  &  &  &  &  &  & \OK & \OK &  &  &  \\  \cdashline{2-7} \cdashline{9-13} \cdashline{15-18} \cdashline{20-23} \cdashline{25-26} \addlinespace[1pt]
\cite{Soltani2019c,kostadinov2021towards} &  &  &  &  &  &  &  &  &  &  &  & \OK &  &  &  &  &  &  &  &  &  &  &  &  &  \\  \cdashline{2-7} \cdashline{9-13} \cdashline{15-18} \cdashline{20-23} \cdashline{25-26} \addlinespace[1pt]
\cite{Linklater2018b} &  &  &  &  &  &  &  &  &  &  & \OK & \OK &  &  &  &  &  &  &  &  &  &  &  &  &  \\  \cdashline{2-7} \cdashline{9-13} \cdashline{15-18} \cdashline{20-23} \cdashline{25-26} \addlinespace[1pt]
\cite{Kim2021a} &  &  &  &  &  &  &  &  &  &  &  & \OK &  &  &  &  &  &  &  &  &  & \OK &  &  &  \\  \cdashline{2-7} \cdashline{9-13} \cdashline{15-18} \cdashline{20-23} \cdashline{25-26} \addlinespace[1pt]
\cite{singh2021private} &  &  &  &  &  &  &  &  &  &  &  & \OK &  &  &  &  &  &  &  & \OK &  &  &  &  &  \\  \cdashline{2-7} \cdashline{9-13} \cdashline{15-18} \cdashline{20-23} \cdashline{25-26} \addlinespace[1pt]
\cite{lockwood2021accessible,WO2021064182A1,Toth2020,shanmugarasa2021towards} &  &  &  &  &  &  &  &  &  &  &  &  &  &  &  &  &  &  &  & \OK &  &  &  &  &  \\  \cdashline{2-7} \cdashline{9-13} \cdashline{15-18} \cdashline{20-23} \cdashline{25-26} \addlinespace[1pt]
\cite{Wohlgemuth2020} &  &  &  &  &  &  &  &  &  &  &  &  &  &  &  &  &  &  &  & \OK &  &  &  &  & \OK \\  \cdashline{2-7} \cdashline{9-13} \cdashline{15-18} \cdashline{20-23} \cdashline{25-26} \addlinespace[1pt]
\cite{inoue2020cooperative,kubach2021lightweight,alber2021adapting} &  &  &  &  &  &  &  &  &  &  &  &  &  &  &  &  &  &  &  &  &  &  &  &  & \OK \\  \cdashline{2-7} \cdashline{9-13} \cdashline{15-18} \cdashline{20-23} \cdashline{25-26} \addlinespace[1pt]
\cite{w3cDID,hardman2020didcomm,DIDExchangeProtocol,smith2019key,fedrecheski2021low} &  &  &  &  &  &  &  &  & \OK &  &  &  &  &  &  &  &  &  &  &  &  &  &  &  &  \\  \cdashline{2-7} \cdashline{9-13} \cdashline{15-18} \cdashline{20-23} \cdashline{25-26} \addlinespace[1pt]
\cite{park2021new} &  &  &  &  &  &  &  &  & \OK &  &  &  &  &  &  &  &  &  &  &  &  & \OK &  &  &  \\  \cdashline{2-7} \cdashline{9-13} \cdashline{15-18} \cdashline{20-23} \cdashline{25-26} \addlinespace[1pt]
\cite{kim2021analysis} &  &  &  &  &  &  &  &  & \OK &  &  &  &  &  &  &  &  &  &  &  & \OK &  &  &  &  \\  \cdashline{2-7} \cdashline{9-13} \cdashline{15-18} \cdashline{20-23} \cdashline{25-26} \addlinespace[1pt]
\cite{naik2021attack} &  &  &  &  &  &  &  &  &  &  &  &  &  &  &  &  &  &  &  &  & \OK &  &  &  &  \\
            \bottomrule
        \end{tabular}}
    \end{adjustwidth}
\end{table}

\subsection{Management}

The \emph{management} facet encompasses five characteristics that deal with the governance of credentials and claims presentation in SSI: (i) \emph{metadata search}; (ii) \emph{protocol integration}; (iii) \emph{identity derivation}; (iv) \emph{wallet security}; and (v) \emph{credential as a service}. These concepts and the works that explore them are presented next.

\subsubsection{Metadata Search}

The authors of~\cite{Lux2019} introduced the problem of \emph{metadata search} in blockchain-based SSI systems. Due to the unstructured nature in which data is stored in blockchain, it becomes a challenge to look for credential metadata stored on the ledger. The authors argued that creating new types of credentials comes at a monetary cost in Sovrin, and thus it is worth reusing existing credential metadata. Hence, effectively tackling the challenge of finding metadata in blockchain-based SSI results in reducing monetary cost for issuers. To attack this problem, the authors of~\cite{Lux2019} used Apache Solr~\cite{solr} to build a search application that allows users to find credential metadata stored in Hyperledger Indy~\cite{indy}, which is the open-source SSI platform that powers Sovrin~\cite{sovrin}. 

Similarly, in~\cite{schardong2021matching} the problem of searching metadata is also explored. The authors employed a natural language processing technique~\cite{spacy} and pre-trained word vectors~\cite{mikolov2013efficient} to enable users to query the Sovrin network's credential metadata using natural language. The reported results outperform~\cite{Lux2019} for queries with synonyms rather than exact terms.

\subsubsection{Protocol Integration}

Another area of study in SSI is \emph{protocol integration} with production-level protocols such as SAML~\cite{hughes2005security}, OAuth 2.0~\cite{hardt2012oauth} and OpenID Connect~\cite{sakimura2014openid}. Failure to successfully address this challenge may jeopardize the adoption of SSI, as billions of users have electronic identities in IdPs that can only communicate using the aforementioned protocols. This challenge was presented as the driving problem in eight research papers~\cite{Gruner2019,Hong2020a,Lux2020,yildiz2021connecting,martinez2021applying,lagutin2019enabling,gruner2021atib} and was also mentioned in three other works~\cite{Schanzenbach2018a,zhong2021jointcloud,Xu2020a}. Three articles aim to integrate SSI with OpenID Connect~\cite{Gruner2019,Lux2020,martinez2021applying}, two works focus on OAuth 2.0~\cite{Hong2020a,lagutin2019enabling}, one on SAML~\cite{lagutin2019enabling}, and one paper on these three protocols~\cite{gruner2021atib}.

Using the OpenID Connect protocol, \cite{Gruner2019} constructs a gateway between two SSI solutions (uPort~\cite{uport} and Jolocom~\cite{jolocom}) and web applications. Users can compose their identities by selecting claims, which are verified by the gateway and then transferred to the destination application for authentication via the OpenID Connect protocol. Similarly, \cite{Lux2020} implements an OpenID Connect gateway between Hyperledger Indy~\cite{indy} and other applications, from which users of any instance of Hyperledger Indy (such as Sovrin~\cite{sovrin}) can benefit. In contrast to~\cite{Gruner2019} a wallet application is designed to store credentials on the user's smartphone. Claims, which the user must present, are used to implement application-level authorization. \cite{martinez2021applying} authenticates the issuer and holder and transfers VCs using OpenID Connect. These VCs include an advanced or qualified signature or seal, which confirms the natural or legal person's identity. A bridge ensures that DID methods and signatures are interoperable among issuers, holders, and verifiers.

Hong \textit{et al.}~\cite{Hong2020a} used OAuth 2.0 for authorization, making it easier to integrate their solution with existing web services. In contrast to~\cite{Gruner2019,Lux2020} authentication in~\cite{Hong2020a} uses a custom mechanism rather than OpenID Connect. Lagutin \textit{et al.}~\cite{lagutin2019enabling} were concerned about the burden of issuing and verifying VPs in resource-constrained devices such as IoT sensors and actuators. A bridge protocol is proposed in which a server receives and processes VPs before distributing modified OAuth 2.0 access tokens to authorized entities. These tokens are given to resource-limited devices, which authorize access to the resource or service.

The authors of~\cite{yildiz2021connecting} proposed an integration with SAML, which allows SSI-based identities to authenticate with SPs via SAML. Gruner \textit{et al.}~\cite{gruner2021atib} presented a more comprehensive architecture that enables users to integrate various SSI offerings with SAML, OpenID Connect, and OAuth 2.0. Additionally, they accomplished \emph{identity derivation}, which is described below, as well as the evaluation of trust models used to accept or deny interactions.

\subsubsection{Identity Derivation}

Allowing users of SSI solutions to access web applications via the OpenID Connect protocol resulted in the implementation of \emph{identity derivation} mechanisms, that is, methods for deriving SSI identities from non-SSI identities. This is the primary goal of~\cite{Abraham2020,abraham2021ssi}, but it was also accomplished in~\cite{Bathen2019,mishra2021pseudo}.

The authors of~\cite{Abraham2020} proposed an electronic identity derivation protocol in which user attributes from various IdPs are collected and transformed into VCs. The transformed VCs can be presented using VPs. Differently, \cite{abraham2021ssi} employs x509 digital certificates~\cite{rfc5280} with high LoA to generate VCs with high LoA. Digital certificates achieve high LoA through a rigorous enrollment process in which the certificate subject must present government-issued documents in person. Both a digital wallet running on a device with a secure enclave and a FIDO2-compatible token~\cite{alliance2019client} equipped with a biometric fingerprint reader generate a key pair after authenticating the owner of an x509 certificate. The VC includes the two public keys. When this VC is used to generate VPs, the private keys of both the digital wallet and the FIDO2 token are accessed. Because the latter requires biometric authentication to perform operations on the private key, the VC holder must be its owner.

Biometric data can be used to make SSI identities, so Bathen \textit{et al.}~\cite{Bathen2019} explored the possibility of replay attacks when an attacker has access to biometric templates. They contended that user-managed cancelable biometrics is the solution to this problem. A person's self-image, \textit{i.e.}, a selfie, is passed through one-way functions to mask the original data, and the resulting data is then stored on a blockchain and managed as a credential. Mishra \textit{et al.}~\cite{mishra2021pseudo} claimed that the underlying techniques used in~\cite{Bathen2019}, namely bloom filters~\cite{patel2015cancelable}, are vulnerable to invertibility and linkability attacks~\cite{hermans2014bloom}. To address these issues, their proposal uses OpenCV~\cite{bradski2000opencv} to extract feature vectors from selfies, which are then subjected to a one-way transformation~\cite{kaur2018random}. Both methods generate revocable biometric credentials suitable for two-factor authentication.

\subsubsection{Wallet Security}

One patent~\cite{WO2021125586A1} is concerned with \emph{wallet security}. Its authors proposed a hardware-based wallet that stores cryptographic keys and credentials. It can connect to mobile devices when necessary and disconnect when not.

\subsubsection{Auditability}

When compared to other identity models, SSI provides more privacy. Nonetheless, some use cases necessitate the \emph{auditability} of credentials or presentations. According to Lemieux \textit{et al.}~\cite{lemieux2021addressing}, there are use cases that require the collection of evidence that a VC was issued and sent to its holder, or that a VP was performed in order to comply with legal, audit, and accountability standards. They proposed using Shamir's Secret Sharing (SSS)~\cite{shamir1979share} to generate a group key capable of encoding and decoding Personal Identifiable Information (PII), such as VCs or VPs, and storing it in a proof registry, \textit{i.e.}, a persistent storage for auditing. This group includes the issuer, the trusted audit service, and the holder. The group key can be generated by two of the three members.

\subsubsection{Credential as a Service}

Three papers discuss the drawbacks of local credential storage and issuance~\cite{samir2021dt,jaroucheh2021secretation,siddiqui2021credentials}. We classify them as credential as a service because their solutions involve outsourcing the storage or processing of credentials.

Samir \textit{et al.}~\cite{samir2021dt} affirmed that storing VCs in a single location is a potential point of failure in SSI implementations because wallets can be lost. Furthermore, they noted that digital wallets confined to a single mobile device might not remain online at all times. To address these concerns, an anonymous multi-party computation solution based on smart contracts and SSS is proposed. It uses SSS to divide a VC into multiple shares, which are then stored on online platforms. Then, smart contracts use multi-party computation to process requests to the VC shares.

In the same way, in~\cite{jaroucheh2021secretation}, holders do not keep their credentials. Credentials are instead stored on a storage service and protected by a two-party protocol. Furthermore, holders do not have direct access to their data. Instead, the VC holder has control over an agent that runs on the storage service and contacts the user to request permission to share information. Users never receive their credentials in this manner, and thus do not have to worry about storing them securely. Because the credentials are encrypted using a two-party encryption protocol, the storage service cannot misuse them.

The authors of~\cite{siddiqui2021credentials} postulated that having the infrastructure to issue credentials is a barrier to SSI adoption. As a result, they proposed using a cloud-based Trusted Execution Environment (TEE)~\cite{sabt2015trusted} to 
issue and distribute VCs to holders.

\subsection{Operational}

The \emph{operational} facet is divided into two facets: \emph{VC} and \emph{VP}. They are a collection of concepts related to the functional aspects of verifiable credentials and verifiable presentations.

\subsubsection{Verifiable Credentials}

\paragraph{Revocation}

Credential \emph{revocation} and status verification are long-standing problems in IAM research. The Online Certificate Status Protocol (OCSP)~\cite{rfc2560} of traditional public key infrastructure (PKI), for example, allows users to query the status of a certificate. However, the query sends the serial number to the Certificate Authority (CA), revealing to the CA where the certificates it issued are being used and thus infringing on user privacy. The revocation 
verification of VCs in a privacy-preserving manner is an active area of research in SSI. Seven works present new approaches to addressing this challenge~\cite{Abraham2020,lax2021lightweight,w3cVC,abraham2020revocable,cho2021verifiable,chotkan2021industry}.

The Verifiable Credentials standard from the World Wide Web Consortium (W3C) defines the meta-structure and lifecycle of VCs and VPs~\cite{w3cVC}. Both VCs and VPs must have the following: (i) metadata describing the data; (ii) the data; and (iii) cryptographic proof of integrity and authenticity. Aside from the roles of issuer, holder, and verifier, a fourth role is the verifiable data registry, which incorporates credential metadata, revocation registries, issuer public keys, and other information. When a model instantiates this metamodel, it must specify the syntax, cryptographic algorithms, and proof format that will be used to construct VCs and VPs. For example, in Hyperledger Indy~\cite{indy}, a VC's metadata is stored in a distributed ledger, whereas the data and proof are stored in a JSON file.

In~\cite{lax2021lightweight}, an approach is detailed in which social media platforms such as Facebook and LinkedIn are used to request, generate, and revoke credentials, as well as present and revoke presented claims. Predicates over credential attributes, on the other hand, are not supported; only attribute disclosure is.

The authors of~\cite{cho2021verifiable} designed a VC that can be issued and revoked by two parties. They argued that this is useful in the financial context. A financial company issues credit scores as VCs together with clients, but these can only be revoked by the financial company with the credit bureau's permission. Their VC includes two digital signatures, one for each entity. A protocol for revocation and status verification using ZKP is proposed.

Chotkan and Pouwelse~\cite{chotkan2021industry} created a mechanism for propagating revocation information using a gossip-based algorithm. Users save the revocation information of their trusted authorities and broadcast it to random peers at predetermined intervals. As a result, issuers are not required to remain online in order to provide revocation data, nor are clients required to contact them in order to obtain such data. The authors provided a threat model as well as a thorough examination of various efficiency metrics.

Abraham \textit{et al.}~\cite{abraham2020revocable} also addressed the issue of offline credential status verification. Their approach is to implement the verifiable data registry as a blockchain, which generates attestation of the validity of requested certificates with a timestamp. When there is no connectivity to the revocation registry, this attestation is presented, and the relying party determines whether it is recent enough to be accepted.

\paragraph{Decentralized Identifiers}

On the internet, entities are identified in a variety of ways. Identification occurs at all levels, from the application to the network. Identifiers are typically issued or controlled by a regulatory agency and assigned to users and machines. IP addresses, for example, are managed by IANA~\cite{iana}, while e-mail providers manage e-mail addresses. A research trend in SSI is to create and improve \emph{decentralized identifiers} from the machine to the human level. Four research articles~\cite{smith2019key,park2021new,kim2021analysis,fedrecheski2021low}, two protocols~\cite{hardman2020didcomm,DIDExchangeProtocol}, and one W3C standard~\cite{w3cDID} have been written in response to various challenges associated with \emph{decentralized identifiers}.

The Decentralized IDentifiers (DID) standard defines a metamodel to create identifiers that are issued and controlled by their owners~\cite{w3cDID}. A DID method is an instance of this metamodel, which sets specific details such as the underlying encryption algorithms and the mechanism by which the method's identifiers are guaranteed to be unique. Each DID is a three-part Uniform Resource Identifier (URI)~\cite{rfc3986} separated by a colon: (i) the \texttt{did} scheme identifier; (ii) the DID method identifier; and (iii) the DID method-specific identifier. For instance, \texttt{did:key:z6MkpTHR8VNsBxYAAWHut2Geadd9jSwuBV8xRoAnwWsdvktH} is a valid DID identifier that uses the DID method \texttt{key}~\cite{did-key-method}. In this method, the first character of the method-specific identifier is always \texttt{z}, and the following three characters represent the public-key algorithm used. In this case, the characters \texttt{6Mk} indicate that \texttt{Ed25519}~\cite{rfc8032} was used, and the subsequent characters are the multibase~\cite{multiformats-multibase-03} encoded public-key. Other DID methods rely on blockchain and other technologies to preserve the user-generated DID and its associated DID document, a JSON-based document with communication endpoints and cryptographic keys to ensure that the holder of a DID is its owner.

Although W3C's DID standard~\cite{w3cDID} provides a foundation for self-sovereign identifiers and the authentication of their owners, it does not define how two (or more) DIDs can interact. The authors of~\cite{hardman2020didcomm} proposed DIDComm, a two-party protocol for establishing a secure communication channel between the holders of two DIDs. It allows messages to be sent via traditional protocols such as HTTP, BlueTooth, NFC, and out-of-band channels such as QRcode and e-mail~\cite{didcomm-out-of-band}. Nonetheless, entities must first exchange DIDs before they can communicate. This is the driving problem of the DID Exchange protocol, which allows DID documents to be exchanged online or offline~\cite{DIDExchangeProtocol}.

According to the authors of~\cite{fedrecheski2021low}, transporting DID documents, which contain identifiers, keys, and communication endpoints, adds a significant overhead to IoT devices. They addressed this issue through three innovations: (i) a new DID method called \texttt{DID:SW} that has a smaller footprint than others; (ii) the use of Concise Binary Object Representation (CBOR)~\cite{rfc7049} to encode DID Documents; and (iii) an extension of DIDComm~\cite{hardman2020didcomm} to DID-based IoT Communication (DIoTComm), which reduces communication parameters and is based on CBOR. The DIoTComm protocol has a five-fold lower overhead than DIDComm.

According to Kim \textit{et al.}~\cite{kim2021analysis}, endpoint URLs in DID documents have an anonymity issue. They claimed that URLs could expose personal information such as country of origin and other affiliations. They proposed two countermeasures: (i) removing URLs and replacing them with other forms of communication; and (ii) using gateway URLs that only redirect authorized entities to the correct address.

From another angle, Smith~\cite{smith2019key} focused on self-certifying identifiers as a means of establishing trust. In this work, user-generated identifiers are coupled to public-key cryptography and explicitly disclose the hash of their next public key in their transactions. This proactive key rotation results in an auditable chain of digital identifier key transfers. To store the history of digital identifiers, a distributed ledger is presented as a root-of-trust.

The key rotation challenge was also addressed in~\cite{park2021new} using Lamport's one-way hash chain~\cite{lamport1981password}. This technique explores the pre-image resistance of cryptographic hash functions by constructing a chain of hash operations on a secret seed and revealing hash values in reverse order. Public-key cryptography is added to this scheme so that only the DID creator, \textit{i.e.}, the person who knows the secret seed, can rotate to the next key pair~\cite{park2021new}.

\paragraph{Issuer Authorization}

Three works present concepts for implementing \emph{issuer authorization}~\cite{schanzenbach2020towards,lauinger2021poa,muktacredential}, which entails issuers creating hierarchies akin to those found in traditional PKI.

Schanzenbach's Ph.D. thesis~\cite{schanzenbach2020towards} describes a structure based on name systems (such as the Domain Name System (DNS)~\cite{rfc882} and the GNU Name System (GNS)~\cite{wachs2013feasibility}) that enables an issuer to delegate authorization to other issuers to issue credentials with specific attributes. Additionally, these secondary issuers have the ability to delegate authorization to other issues, and so on.

With the same objective in mind, but a different approach, the authors of~\cite{lauinger2021poa} formalized a model that utilizes the RSA cryptographic accumulator~\cite{camenisch2002dynamic} to enable authorized issuers to issue credentials without disclosing their identity. The authors argued that this addresses a gap in the Hyperledger Indy framework~\cite{indy}, in which an issuer $A$ cannot prevent another issuer $B$ from issuing credentials in the same format as $A$.

According to the authors of~\cite{muktacredential}, VCs issued in SSI today are assumed to be from trusted issuers, such as government agencies. Their work proposes an issuer authorization scheme based on policies, in which an issuer is only authorized to issue VCs if its policy allows it to. The root of authority serves as the policy authority, defining policies for issuers.

\paragraph{Delegation}

Three research papers propose methods for achieving credential \emph{delegation}. It refers to an individual's or group's ability to delegate some of their identity data to another individual or group of individuals. Two of them~\cite{Linklater2018b,Liu} are discussed later in this manuscript (Section~\ref{recovery} and Section~\ref{architecture}), as \emph{delegation} is not their primary goal.

Lim \textit{et al.}~\cite{lim2021subject} proposed a system for VC delegation that requires the VC subject to confirm or deny the delegatee's use of the VC. A VP constructed by delegatees is limited in their method, as they only have the VC in an encrypted format. As a result, any VP presented by a delegatee induces communication with the VC subject in order to obtain authorization and incorporate the VP with required data.

\paragraph{Backup and Recovery}\label{recovery}

Another trend of research in SSI is the \emph{backup and recovery} of keys and certificates. Empowering users with the ability to control their credentials currently comes with many burdens that were previously the tasks of IdPs. At this point, the \emph{backup and recovery} of identity-associated materials are significant burdens. Proposing backup and recovery mechanisms to keys and credentials are the main objective of six research papers~\cite{Soltani2019c,kostadinov2021towards,Linklater2018b,Jakubeit2020,Kim2021a,singh2021private}.

Soltani \textit{et al.}~\cite{Soltani2019c} used a decentralized protocol to handle key recovery. They created a wallet application in which users define their trusted peers and the recoverable keys. In a protocol based on SSS~\cite{shamir1979share}, key pieces are distributed to trusted users and can be recovered by the owner if a minimum number of parts can be retrieved from peers.

The authors of~\cite{singh2021private} presented a trade-off between security (storing an encrypted form of the private key in lower security environments) and usability (recovering the original private key without the need for long passwords or Hardware Security Modules (HSMs)). The private key is divided using SSS~\cite{shamir1979share} to achieve this trade-off. The user must correctly answer a minimum number of previously registered questions, with each response constituting a component of SSS. To improve security, the minimum number of correct answers might be increased.

\cite{kostadinov2021towards} also addresses the issue of identity recovery. Its authors suggested that a suitable solution would be to use another device in the identity owner's possession as a storage provider. To improve usability, it has been recommended that protocols could be developed and integrated with routers, resulting in a seamless user experience.

In~\cite{Linklater2018b}, a self-signed root certificate acts as a CA that creates short-lived certificates for the users. The authors concluded that because certificates are rotated on a predetermined schedule, the key recovery issue is resolved as long as the 
CA's private key remain intact.

A private data backup system with two additional roles is proposed in~\cite{Jakubeit2020}: trusted audit service and trusted individuals. The trusted audit service receive portions of the keys. In contrast, the trusted individuals must physically meet to receive encrypted shares of the private data to store on short-range connectivity devices (such as infrared or near-field communication). Following the loss of personal data, trusted peers meet and confirm the affected user's newly generated electronic identity to the trusted audit service, which provide the user with the key necessary to decrypt the private data gathered from trusted peers.

From a different perspective, \cite{Kim2021a} uses proxy re-encryption~\cite{blaze1998divertible}. This technique allows data encrypted with a person's key to be decrypted using someone else's key without revealing anyone's data or key to the proxy. Trusted individuals execute a group key agreement, and then the derived group key is sent to the proxy that contains the encrypted user data. The user's private data can be retrieved from the proxy if the group recreates its key and uses it to authenticate with the proxy, which then uses the proxy re-encryption scheme to have the user's private data accessible to the group.

\subsubsection{Verifiable Presentation}

\paragraph{Revocation}\label{revocation}

A challenging topic in SSI research is the \emph{revocation} of VPs. Four research endeavors aspire to solve it~\cite{Stokkink2018c,lax2021lightweight,Xu2020a,Schanzenbach2018a}, one of which was presented above~\cite{lax2021lightweight}.

Concerned about the portability and interoperability of VPs, the authors of~\cite{Stokkink2018c} introduced a metamodel for specifying VPs in blockchains. The VP metadata consists of the following elements: name, timestamp, expiration time, proof format, and proof link. The VP lifecycle is structured in a blockchain format with two types of blocks: one for adding a signature to a VP and one for revoking a signature. If all of the signatures endorsing a VP are revoked, the VP is deemed revoked as well.

The authors of \cite{Xu2020a}, on the other hand, used chameleon hashing~\cite{krawczyk2000chameleon} to implement VP revocation. This one-way function family employs a trapdoor, so that without it, they behave similarly to traditional one-way functions. If one has access to the trapdoor, such as via a key, one can easily find collisions for a given input. This special feature was used in \cite{ateniese2017redactable,cryptoeprint:2019:406} to implement a rewriteable blockchain, that is, a blockchain whose history can be manipulated via the chameleon hash trapdoor. Based on these efforts, the authors of~\cite{Xu2020a} designed their blockchain to allow users to revoke access to VPs anchored in the ledger via a trapdoor.

Lastly, in~\cite{Schanzenbach2018a}, it is argued that VPs cannot be revoked because they are likely to have been persisted locally by the RPs. The proposed solution to this problem is to grant access to up-to-date information via version control and encryption. Keys are distributed to authorized RPs. 

\paragraph{Verifier Authorization}

\emph{Verifier authorization} is a relatively new topic. The idea is to give issuers some control over the credentials they issue by establishing rules that verifiers must follow in order to access holders' VPs. This appears to conflict with the philosophical basis of SSI, which specifies that issuers should not dictate what holders of VCs may or may not do.

The authors of~\cite{anaigoundanpudur2021cryptographic} used Ciphertext-Policy Attribute-Based Encryption (CP-ABE)~\cite{bethencourt2007ciphertext} to allow issuers to create a policy imposing minimum requirements on verifiers requesting VPs from holders. The decryption key in CP-ABE is derived from the attributes of the deciphering entity. A doctor, for example, who receives patient data must have a doctor registration VC and be specialized in a particular field.

\paragraph{Data Minimization}

Perhaps the most valuable feature of SSI for individuals is its emphasis on \emph{data minimization}. Three types of techniques are described in the literature: (i) selective disclosure~\cite{Abraham2020,abraham2020revocable,yang2020zero}, which enables the creation of VPs containing only some of the attributes of a VC rather than all of them; (ii) predicates, \textit{i.e.}, boolean assertions over data~\cite{bobolz2021issuer}; and (iii) arbitrary statements over attributes~\cite{lee2021privacy,Lee2019,Schanzenbach2019}.

Abraham \textit{et al.}~\cite{Abraham2020} built a ZKP proof system using Water's signature~\cite{waters2005efficient} and BLS signature~\cite{boneh2001short} that enables selective disclosure of certificate attributes. The same technique is employed in~\cite{abraham2020revocable}. Similarly, \cite{yang2020zero} uses zero-knowledge Succinct Non-interactive ARguments of Knowledge (zk-SNARK)~\cite{bitansky2012extractable} to create a VP format where holders can prove possession of a specific attribute and reveal its value. 

In~\cite{bobolz2021issuer}, ZKP allows the creation of a VP to mathematically prove that a VC was created by an issuer who is a member of a group of authorized issuers without revealing any unique identifier, such as the issuer's public key. Finally, the authors of~\cite{lee2021privacy}, \cite{Lee2019} and~\cite{Schanzenbach2019} enable credential holders to explore the full expressive power of zk-SNARK, \textit{i.e.}, to produce proofs in any language in \texttt{NP}.

\paragraph{Reuse Prevention}

Nothing stops the RP from copying what it learns from the user after receiving a VP. \emph{Preventing the reuse} of acquired knowledge is one of the most challenging aspects of SSI.

The creators of~\cite{kang2021decentralized} attempt to solve this challenge. They proposed an architecture that allows holders to charge RPs to access their attributes while preventing reuse. Instead of selective disclosure or proofs over private data, Fully Homomorphic Encryption (FHE)~\cite{gentry2009fully} is used. FHE is a method for processing encrypted data and producing valid results without decryption. Their proposal uses FHE to process user data in a secure third-party environment that both the user and the RP trust. According to the authors, this technique prevents private information from being leaked.
Although it is unlikely that FHE will reveal user attributes, information about the computation over private data can be revealed.

\subsection{System Design}

The facet \emph{system design} encompasses four concepts related to the conceptualization of SSI: \emph{design/architecture}, \emph{Human-Computer Interactions (HCI)}, \emph{risk assessment} and \emph{security model}.

\subsubsection{SSI Design/Architecture}\label{architecture}

Five articles discuss various aspects of what we refer to as \emph{SSI Design} or \emph{SSI Architecture}~\cite{w3cVC,Stokkink2018c,Liu,Barclay2020a,Liu2020,Ferdous2019}. Rather than addressing specific issues or proposing SSI systems, these publications explore and analyze the planning, design, and construction of SSI systems. Previously, the W3C's VC metamodel~\cite{w3cVC} and Stokkink \textit{et al.}'s~\cite{Stokkink2018c} VP metamodel were examined. This section discusses the remaining three research papers in this category.

In~\cite{Liu}, design patterns are presented to assist in the development of new SSI applications on the blockchain. The lifecycles of key management, identity management, and credential management are discussed. Then, twelve patterns are proposed within these three groups, following Martin \textit{et al.}'s~\cite{martin1997pattern} format, which includes a pattern name, summary, context of use, problem statement, discussion, solution, and its consequences.

On the other hand, the authors of~\cite{Liu2020} asserted that identity management systems could be reduced to two mappings: (i) digital identifier and its owner, and (ii) digital identifier and its credentials. Furthermore, for both mappings, the following operations are required: create, read, update, delete, and verify. The system's chosen trust model determines the manner in which they are built. If the goal is SSI, all of them should be completed independently of any authority.

Barclay and colleagues~\cite{Barclay2020a} demonstrated a modeling technique that enables non-technical stakeholders to specify and comprehend SSI entities and their relationships. They used iStar 2.0~\cite{dalpiaz2016istar}, an actor-based modeling language that enables the representation of actors and the interdependence of their goals. In an SSI system, the actors are the users who issue credentials and present claims.

Finally, Ferdous \textit{et al.}~\cite{Ferdous2019} created a detailed mathematical model of SSI. This formalization includes a feature that is unique in the SSI literature reviewed: user de-registration.

\subsubsection{HCI}

There are five research materials~\cite{Toth2020,shanmugarasa2021towards,lockwood2021accessible,Wohlgemuth2020,singh2021private} and one patent~\cite{WO2021064182A1} that look into usability and human perception issues in SSI systems. Section~\ref{recovery} already introduced the work of Sign \textit{et al.}~\cite{singh2021private}. They are grouped under the \emph{HCI} concept of our taxonomy.

Toth \textit{et al.}~\cite{Toth2020} claimed that biometrics and other forms of two-factor authentication only marginally improved identity security. They then introduced a software agent to manage user data. It helps users decide which credentials to use and which private information to reveal, improving security through improved human-computer interactions.

With a different emphasis, the authors of~\cite{WO2021064182A1} submitted a patent for an authentication method based on a users' interactions with their personal device. To determine if the person holding the device is the owner, the device monitors application usage patterns, browser history, location history, and other measurements.

Pertaining HCI and trust, \cite{Wohlgemuth2020} suggest that deciding whether or not to trust an identity and its claims is a major risk for an algorithm to decide on its own. The authors put forward a proposal in which the user must actively decide whether electronic identities can be trusted. The user is empowered to make that decision by viewing a graph of the proponent's previous interactions with other electronic identities, which is generated from the history stored in a distributed ledger.

The authors of~\cite{lockwood2021accessible} presented an extensive study of SSI usability and discovered that current SSI systems interactions necessitate extensive prior knowledge and participant responsibility. The authors investigated the SSI interface layer using the human data interaction theory~\cite{mortier2015humandata}, which says that humans interact with data rather than computers. To increase the likelihood of adoption, the conclusion emphasizes the need for standardization and design thinking of interfaces and interactions.

Shanmugarasa \textit{et al.}~\cite{shanmugarasa2021towards} addressed the issue of users managing VPs. Non-technically competent users, for instance, may agree to submit more information than the RPs actually needs. The proposed solution to this problem is a privacy preference recommendation system that employs machine learning algorithms and pre-trained models based on survey data on privacy preferences. This system assists the user by suggesting on which attributes can be shared.

\subsubsection{Risk Assessment and Threat/Attack Model}

In relation to the design of SSI, two concepts related to computer security were observed in the reviewed literature, namely \emph{risk assessment} and \emph{threat/attack model}. The latter entails two activities: (i) identifying and analyzing potential threats; and (ii) comprehending how an attacker can exploit them. These two tasks are part of the risk assessment, which also includes calculating the potential loss if a vulnerability is exploited. Eighteen works described in the other sections incorporated one or both of these activities to improve their schemes. While three articles discussed risk assessment, only one makes a novel contribution by tying risk assessment and SSI together~\cite{naik2021attack}.

Naik \textit{et al.}~\cite{naik2021attack} developed a tree-based risk analysis method for SSI. The attack tree approach represents the attack goal as the root of a tree, and the methods and actions to achieve the goal as the leaves~\cite{schneier1999attack}. In this work, important assets in an SSI system are identified first. Then, the attack tree is used to generate input for their risk analysis, which concludes with appropriate mitigations for the identified risks.

\subsection{Trust}

The final practical facet of our taxonomy is \emph{trust}. Entities in any IAM model must decide whether they trust other entities and, as a result, the data they generate. Since the inception of SSI, a strong emphasis has been placed on the use of verifiable credentials in order for RPs to be certain about the origin of the credentials~\cite{w3cVC}.

SSI promotes the decentralization of identity management. Furthermore, the majority of SSI offerings endorse the deconstruction of centralized sources of trust (\textit{e.g.} IANA~\cite{iana} and Certification Authority Browser Forum~\cite{caforum}). Most SSI platforms allow anyone to issue VCs in anyone's name. As a result, \emph{reputation models} that allow RPs to quantitatively assess whether a VP (and thus a VC) is trustworthy or not have been an active topic of study. Another topic of interest is the development of \emph{trust policy evaluation} techniques for evaluating policy-based reputation models.

\subsubsection{Reputation Model}

Six research articles present or discuss \emph{reputation models} for SSI~\cite{Gruner2018,Gruner2019a, bhattacharya2020enhancing,abramson2021evaluating,zhong2021jointcloud,Stokkink2018c}. Section~\ref{revocation} introduced one of them. The rest are described below.

Gruner \textit{et al.}~\cite{Gruner2018} used graph theory to model trust in blockchain-based SSI systems. The originator of VPs is endorsed in a blockchain by system participants in their proposal. This enables the creation of an endorsement graph. They proposed an algorithm that navigates the graph and calculates a trust factor for the system's participants. This trust factor can be used to determine whether a participant can be trusted or if they are a malicious user.

Bhattacharya \textit{et al.}~\cite{bhattacharya2020enhancing} expanded on~\cite{Gruner2018} by including time as a variable in their reputation model. They hypothesized that, in the context of Sovrin, the initial reputation of issuers could be influenced by Sovrin's onboarding process, which could be biased or falsified.

The authors of~\cite{Gruner2019a}, on the other hand, developed a probabilistic model of trust. They applied probability theory to determine whether claims about the same information from different issuers could be combined to generate trust about it.

Zhong \textit{et al.}~\cite{zhong2021jointcloud} raised the problem of current SSI offerings' lack of interoperability and how this restricts the evaluation of VC credibility. Their solution to this problem employs cross-chain smart contracts to compute a credibility score based on the boolean evaluation (either support or refuse) of all verifiers who verify the VC, taking into account each verifier's credibility.

Finally, Abramson \textit{et al.}~\cite{abramson2021evaluating} described the different user roles and transaction types stored in the Hyperledger Indy blockchain, including the steps a verifier can take to gain confidence when receiving a presentation. For example, they argued that if multiple entities issue credentials of a given format (credential schema), this provides more assurance than a schema that is only endorsed by a single issuer.

\subsubsection{Trust Policy Evaluation}

The \emph{trust policy evaluation} is covered in eight papers~\cite{Gruner2019,gruner2021atib,martinez2021applying,lauinger2021poa,Wohlgemuth2020}. Three of them~\cite{Gruner2019,gruner2021atib,martinez2021applying}, which were previously introduced, are concerned with \emph{protocol integration} and \emph{identity derivation}. One aims for \emph{issuer authorization}~\cite{lauinger2021poa}, while the other for \emph{HCI}~\cite{Wohlgemuth2020}. The following are the three papers that attempt to address this problem~\cite{inoue2020cooperative,kubach2021lightweight,alber2021adapting}.

The authors of~\cite{kubach2021lightweight} proposed that entities define trust policies through lists of authorities they trust. These trusted entities, in turn, also publish which entities they recognize as trustworthy. For instance, one could trust a bank federation that periodically reports which banks it recognizes as credible. Thus, when receiving the VP of a person stating that she has an account on an unrecognized bank, a query to the bank federation's list of trusted banks is enough to decide if the VP can be trusted or not. 

Inoue \textit{et al.}~\cite{inoue2020cooperative} considered the task of updating an individual's information across multiple issuers and RPs, each with its own trust policy. This challenge was modeled as an Integer Linear Programming (ILP) problem, with trust policies defined as credibility requirements for incoming update requests. Updating a person's information in an issuer or RP increases its credibility. The ILP is then transformed into a graph problem, and an approximate solution is found using a heuristic based on Dijkstra's algorithm. This article is the only one in the survey that provides a formal description of the problem.

The Trust Policy Language (TPL)~\cite{modersheim2019}, a declarative language for specifying trust rules without concern for low-level details, was adapted to work in SSI in~\cite{alber2021adapting}. The TPL has been enhanced with SSI-related concepts such as DID and VC, allowing the specification of rules to validate VPs.

\section{RQ-2: What properties, formal definitions and cryptographic tools have been used?}\label{sec:RQ2}
The first two years of examined papers were mostly focused on conceptual contributions to SSI. From 2018 forward, the works evaluated began to provide mathematical structures to help properly represent concepts. There are twenty-seven articles in total that include some type of formalism. Table~\ref{tbl:R2} shows these articles and the building blocks they utilized.
We divide the formal definitions into two categories: cryptographic tools and non-cryptographic tools. Cryptographic tools are well-known, low-level cryptographic algorithms that are often employed in computer systems to develop secure protocols and systems~\cite{menezes2018handbook,barker2016guideline}. 

\begin{table}[h]
\begin{adjustwidth}{-0.55cm}{0cm}
    \caption{Publications that introduce mathematical formalism to SSI. Techniques are divided into cryptographic and non-cryptographic tools.}
    \label{tbl:R2}
    \begin{tabular}{lrcccccccccccccccc}
        \toprule
         \multirow{2}{*}[-1.2em]{Concept} & \multirow{2}{*}[-1.2em]{Works} & \multirow{2}{*}{\rot{Formalism}} && \multicolumn{4}{c}{\begin{tabular}[c]{@{}c@{}}Non-Crypto.\\ Tools\end{tabular}} && \multicolumn{9}{c}{Cryptographic Tools} \\ \cmidrule{5-8} \cmidrule{10-18}
         &&&& \rot{ILP} & \rot{Graph} & \rot{NS} & \rot{Prob.} && \rot{MPC} & \rot{SSS} & \rot{PRE} & \rot{CH} & \rot{ABE} & \rot{ZKP} & \rot{CAcc} & \rot{MS} & \rot{FHE} \\ \midrule
        Identity Derivation & \cite{abraham2021ssi} & \OK &  &  &  &  &  &  &  &  &  &  &  &  & \OK &  & \\ \hdashline \addlinespace[1pt]
        Credential as a Service & \cite{samir2021dt} & \OK &  &  &  &  &  &  & \OK & \OK &  &  &  &  &  &  &  \\ \hdashline \addlinespace[1pt]
        \multirow{5}{*}{Revocation} & \cite{Schanzenbach2018a} & \OK &  &  &  & \OK &  &  &  &  &  &  & \OK &  &  &  &  \\
        & \cite{Xu2020a} & \OK &  &  &  &  &  &  &  &  &  & \OK &  &  &  &  &  \\
        & \cite{cho2021verifiable} & \OK &  &  &  &  &  &  &  &  &  &  &  & \OK &  &  &  \\
        & \cite{lax2021lightweight,chotkan2021industry} & \OK &  &  &  &  &  &  &  &  &  &  &  &  &  &  &  \\ 
        & \cite{abraham2020revocable} & \OK &  &  &  &  &  &  &  &  &  &  &  & \OK &  & \OK &  \\ \hdashline \addlinespace[1pt]
        Decentralized Identifiers & \cite{smith2019key} & \OK &  &  &  &  &  &  &  &  &  &  &  &  &  &  & \\ \hdashline \addlinespace[1pt]
        \multirow{2}{*}{Issuer Authorization} & \cite{lauinger2021poa} & \OK &  &  &  &  &  &  &  &  &  &  &  & \OK & \OK &  & \\
        & \cite{schanzenbach2020towards} & \OK &  &  &  & \OK &  &  &  &  &  &  & \OK &  &  &  &  \\ \hdashline \addlinespace[1pt]
        \multirow{3}{*}{Backup and Recovery} & \cite{Jakubeit2020} & \OK &  &  &  &  & \OK &  &  &  &  &  &  &  &  &  & \\
        & \cite{Soltani2019c,singh2021private} & \OK &  &  &  &  &  &  &  & \OK &  &  &  &  &  &  &  \\
        & \cite{Kim2021a} & \OK &  &  &  &  &  &  &  &  & \OK &  &  &  &  &  &  \\ \hdashline \addlinespace[1pt]
        Verifier Authorization & \cite{anaigoundanpudur2021cryptographic} & \OK &  &  &  &  &  &  &  &  &  &  & \OK &  &  &  &  \\ \hdashline \addlinespace[1pt]
        \multirow{2}{*}{Data Minimization} & \cite{Lee2019,Schanzenbach2019,yang2020zero,lee2021privacy} & \OK &  &  &  &  &  &  &  &  &  &  &  & \OK &  &  & \\
        & \cite{Abraham2020} & \OK &  &  &  &  &  &  &  &  &  &  &  & \OK &  & \OK &  \\ \hdashline \addlinespace[1pt]
        Reuse Prevention & \cite{kang2021decentralized} & \OK &  &  &  &  &  &  &  &  &  &  &  &  &  &  & \OK \\ \hdashline \addlinespace[1pt]
        SSI Design/Architecture & \cite{Ferdous2019} & \OK &  &  &  &  &  &  &  &  &  &  &  &  &  &  &  \\ \hdashline \addlinespace[1pt]
        \multirow{2}{*}{Reputation Model} & \cite{Gruner2019a} & \OK &  &  &  &  & \OK &  &  &  &  &  &  &  &  &  &  \\
        & \cite{Gruner2018,bhattacharya2020enhancing} & \OK &  &  & \OK &  &  &  &  &  &  &  &  &  &  &  &  \\ \hdashline \addlinespace[1pt]
        Trust Policy Evaluation & \cite{inoue2020cooperative} & \OK &  & \OK & \OK &  &  &  &  &  &  &  &  &  &  &  & \\ 
        \bottomrule
    \end{tabular}
\end{adjustwidth}
\end{table}

Inoue \textit{et al.}~\cite{inoue2020cooperative} modeled trust policy evaluation using \emph{Integer Linear Programming (ILP)}. ILP is a mathematical optimization formulation in which all variables are integers and the objective function is linear~\cite{schrijver1998theory}. It may be used with other formulations, such as \emph{graph} theory to express graph-related problems. For instance, the shortest path between two nodes. In addition to~\cite{inoue2020cooperative}, two other papers used graph models to generate reputation models~\cite{Gruner2018,bhattacharya2020enhancing}.

Two works lead by Martin Schanzenbach~\cite{Schanzenbach2018a,schanzenbach2020towards} used \emph{Name System (NS)} (\textit{e.g.} Domain Name System (DNS)~\cite{rfc882}, and GNU Name System (GNS)~\cite{wachs2013feasibility}) as blocks for attacking revocation and issuer authorization challenges. These systems are coupled with \emph{Attribute-Based Encryption (ABE)}, which allow the user to selectively give and revoke access to some of their attributes to reach their objectives. Another work that models a solution based on ABE is~\cite{anaigoundanpudur2021cryptographic}.

Lastly in the non-cryptographyc tools category is \emph{probability theory}. Both Gruner \textit{et al.}~\cite{Gruner2019a} and Jakubeit \textit{et al.}~\cite{Jakubeit2020} base their contributions on this branch of mathematics.

We mapped nine cryptographic techniques formally defined in the examined literature. Most of the practical research we surveyed discussed about how cryptographic primitives like public-key cryptography and hash functions are used. Nevertheless, we only included those that did so with more than simple textual explanations in this study.

\emph{Multi-Party Computation (MPC)} is formally described and used in~\cite{samir2021dt}. This field of research investigates methods for parties to compute a function together over their inputs without revealing them to the other parties~\cite{goldreich1998secure}. In~\cite{samir2021dt}, MPC was used in conjunction with \emph{Shamir's Secret Sharing (SSS)}~\cite{shamir1979share}. This technique was used in two other articles to achieve backup and recovery of credentials~\cite{Soltani2019c,singh2021private}. The SSS algorithm breaks a secret into shares. The original secret is recalculated using a predetermined number of shares, generally fewer than the total number of shares.

Another technique that was precisely described in the SSI literature was \emph{Proxy Re-Encryption (PRE)}~\cite{blaze1998divertible}. This technique allows data encrypted with a person's key to be decrypted using someone else's key without revealing anyone's data or key to the proxy. It was used by Kim \textit{et al.}~\cite{Kim2021a} to recover private data.

The authors of~\cite{Xu2020a} implemented VP revocation with \emph{Chameleon Hashing (CH)}. This family of one-way functions employs a trapdoor to find collisions for a given input~\cite{krawczyk2000chameleon}.

User privacy is the utmost goal of SSI, and the most popular technique used to increase privacy is to use \emph{Zero-Knowledge Proof (ZKP)} to convince the RP of statements regarding the user's private information. Five articles that mainly propose data minimization techniques formally defined their approaches~\cite{Lee2019,Schanzenbach2019,yang2020zero,lee2021privacy,Abraham2020}, four of which use zk-SNARK to achieve ZKP~\cite{yang2020zero,lee2021privacy,Lee2019,Schanzenbach2019} and the other~\cite{Abraham2020} uses \emph{Multi-Signature (MS)}, which is also employed in~\cite{abraham2020revocable}. MS allows a set of participants to sign a document or message. 
Two papers formally describe and use \emph{Cryptographic Accumulator (CAcc)} as part of their solutions~\cite{abraham2021ssi,lauinger2021poa}. CAcc is a data structure that enables the accumulation of a large set of values into one short accumulator. One of the characteristics of CAcc is that values can be added and set membership verified in constant time. The authors of~\cite{abraham2021ssi} use it as part of the process of creating SSI identities from traditional PKI-based identities and~\cite{lauinger2021poa} to achieve issuer authorization. 

Lastly, \emph{Fully Homomorphic Encryption (FHE)}~\cite{gentry2009fully} is used to prevent reuse of presented information in~\cite{kang2021decentralized}. FHE allows encrypted data to be processed without decryption. 

\section{RQ-3: What conceptual ideas have been introduced or refuted?}\label{sec:RQ3}
Christopher Allen~\cite{Path2016} stated that there is currently no agreement on a definition of SSI and then presented ten guiding principles as a starting point. Our third research question is answered by an examination of the literature's debates on the SSI definition, which is now presented to the reader.

We found seventeen works that contribute to Allen's discussion regarding the meaning of SSI by using our review process. Table~\ref{tbl:R3} summarizes these studies in accordance with our taxonomy, which has the facets \emph{add} and \emph{refute} under \emph{conceptual}. Furthermore, the facet \emph{add} is subdivided into \emph{functional} and \emph{non-functional}.

\begin{table}[h]
\begin{adjustwidth}{-\extralength}{0cm}
\caption{Publications that add or refute philosophical views of SSI.}
\label{tbl:R3}
\resizebox*{\linewidth}{!}{
\begin{tabular}{rccccccccccccccccccccccccc}
    \toprule
    & \multicolumn{17}{c}{Add} && \multicolumn{7}{c}{\multirow{2}{*}[-0.25em]{Refute}} \\ \cmidrule{2-18}
    & \multicolumn{8}{c}{Functional} && \multicolumn{8}{c}{Non-Functional} && &&&&&& \\
    \cmidrule{2-9} \cmidrule{11-18} \cmidrule{20-26}
    Works & \rot{No Central Authority} & \rot{Legacy Compatible} & \rot{Verifiable Presentation} & \rot{Counterfeit Prevention} & \rot{Identity Verification} & \rot{Identity Assurance} & \rot{Secure transactions} & \rot{Delegation} &  & \rot{Recoverability} & \rot{Usability} & \rot{Accessibility} & \rot{Availability} & \rot{Auditability} & \rot{Scalability} & \rot{Free} & \rot{Regulatory} &  & \rot{Existence} & \rot{Control} & \rot{Access} & \rot{Consent} & \rot{Persistence} & \rot{Transparency} & \rot{Protection} \\ \midrule
\cite{Stokkink2018c,Muhle,VCfaq2017} &  &  & \OK &  &  &  &  &  &  &  &  &  &  &  &  &  &  &  &  &  &  &  &  &  &  \\ \cdashline{2-9} \cdashline{11-18} \cdashline{20-26} \addlinespace[1pt]
\cite{Andrieu2016} & \OK & \OK & \OK &  &  &  &  &  &  & \OK &  &  &  &  & \OK & \OK &  &  &  &  &  &  &  &  &  \\ \cdashline{2-9} \cdashline{11-18} \cdashline{20-26} \addlinespace[1pt]
\cite{Naik2020a} &  &  & \OK &  &  &  & \OK &  &  & \OK &  & \OK & \OK &  & \OK & \OK &  &  &  &  &  &  &  &  &  \\ \cdashline{2-9} \cdashline{11-18} \cdashline{20-26} \addlinespace[1pt]
\cite{Ferdous2019} &  &  &  &  &  &  & &  &  & &  & & \OK &  &  & &  &  &  &  &  &  &  &  &  \\ \cdashline{2-9} \cdashline{11-18} \cdashline{20-26} \addlinespace[1pt]
\cite{Ellingsen2019a} &  &  &  &  &  &  &  &  &  &  &  &  &  &  &  & \OK &  &  & \OK &  &  &  &  &  &  \\ \cdashline{2-9} \cdashline{11-18} \cdashline{20-26} \addlinespace[1pt]
\cite{VanWingerde2017a} &  &  & \OK &  &  &  &  &  &  & \OK &  &  &  &  &  &  & \OK &  &  &  &  &  &  &  &  \\ \cdashline{2-9} \cdashline{11-18} \cdashline{20-26} \addlinespace[1pt]
\cite{Abraham2017} & \OK &  &  &  &  &  &  &  &  &  &  &  &  &  &  &  &  &  &  &  &  &  &  &  &  \\ \cdashline{2-9} \cdashline{11-18} \cdashline{20-26} \addlinespace[1pt]
\cite{Satybaldy2020d} & \OK &  &  &  &  &  &  &  &  &  & \OK &  &  &  & \OK &  &  &  &  &  &  &  &  &  &  \\ \cdashline{2-9} \cdashline{11-18} \cdashline{20-26} \addlinespace[1pt]
\cite{Diebold2017a} & \OK &  &  &  &  &  &  &  &  & \OK & \OK &  &  &  &  &  &  &  &  &  &  &  &  &  &  \\ \cdashline{2-9} \cdashline{11-18} \cdashline{20-26} \addlinespace[1pt]
\cite{Speelman2020a} &  &  &  &  &  &  &  &  &  &  & \OK &  &  &  &  &  &  &  &  &  &  &  &  &  &  \\ \cdashline{2-9} \cdashline{11-18} \cdashline{20-26} \addlinespace[1pt]
\cite{Toth} &  &  &  & \OK & \OK & \OK & \OK &  &  &  & \OK &  &  &  &  &  &  &  & \OK &  &  &  &  & \OK & \OK \\ \cdashline{2-9} \cdashline{11-18} \cdashline{20-26} \addlinespace[1pt]
\cite{chotkan2021industry} &  &  & \OK &  &  &  &  &  &  &  &  &  &  &  &  &  & \OK &  &  &  &  &  &  &  &  \\ \cdashline{2-9} \cdashline{11-18} \cdashline{20-26} \addlinespace[1pt]
\cite{vcuvcko2021towards} & \OK & \OK &  &  &  & \OK & \OK &  &  & \OK & \OK & \OK &  &  &  &  &  &  &  &  &  &  &  &  &  \\ \cdashline{2-9} \cdashline{11-18} \cdashline{20-26} \addlinespace[1pt]
\cite{sovrin2021principles} & \OK &  & \OK &  &  &  & \OK & \OK &  &  & \OK & \OK &  &  &  &  &  &  &  &  &  &  &  &  &  \\ \cdashline{2-9} \cdashline{11-18} \cdashline{20-26} \addlinespace[1pt]
\cite{Schutte2016} &  &  &  &  &  &  &  &  &  &  &  &  &  & \OK &  &  &  &  & \OK & \OK & \OK & \OK & \OK &  &   \\ 
    \bottomrule
\end{tabular}}
\end{adjustwidth}
\end{table}

\subsection{Add}

\subsubsection{Functional}

\emph{No central authority} means that no single organization should be in charge of or own an SSI solution~\cite{Andrieu2016,Abraham2017,Satybaldy2020d,Diebold2017a,vcuvcko2021towards,sovrin2021principles}. The articles that define this property, as well as the articles that say that SSI should be \emph{free}~\cite{Andrieu2016,Ellingsen2019a,Naik2020a}, make good arguments at first glance. However, upon closer examination, these characteristics may discourage businesses from investing in SSI. They would have to seek alternative sources of income and share control over their products. To some extent, this is what Evernym~\cite{evernym}, a for-profit company, did when it split off Sovrin, a non-profit foundation that is supported by other organizations~\cite{tobin2016inevitable}. Sovrin, on the other hand, is not free. While end users can join the network, receive VCs, and issue VPs for free, companies or other entities that enroll their end users must pay fees to~\cite{sovrinIssueCredentials}: (i) join the network; (ii) register a credential format, i.e., a credential schema; (iii) begin issuing credentials using a registered schema; (iv) register a revocation registry; and (v) revoke VCs.

According to two studies, SSI systems must be \emph{compatible with legacy} identity management systems and protocols~\cite{Andrieu2016,vcuvcko2021towards}. According to the reviewed literature, this is a highly researched subject. The applied research focuses on two aspects of legacy compatibility: (i) protocol integration with prior standards such as SAML, OAuth 2.0, and OpenID Connect; and (ii) identity derivation in order to migrate identities from identity providers that adopt the aforementioned protocols to SSI systems.

According to~\cite{Stokkink2018c,Muhle,VCfaq2017,Andrieu2016,Naik2020a,VanWingerde2017a,chotkan2021industry,sovrin2021principles}, the concept of \emph{verifiable presentation} is an integral part of SSI such that, without it, we cannot achieve SSI. 

Toth and Anderson-Priddy~\cite{Toth} defined four additional \emph{functional} properties of SSI, two of which have not been accounted for by others: (i) \emph{counterfeit prevention}, which involves the impossibility of producing fake identities from others; and (ii) \emph{identity verification}, which requires interacting parties to be assured of the authenticity of the identity owner. According to the property \emph{identity assurance}, which has been proposed elsewhere~\cite{vcuvcko2021towards}, entities that rely on (self-sovereign) identities should be able to see proof that the entities with whom they interact are who they claim to be. The fourth additional property proposed by~\cite{Toth} and others~\cite{Naik2020a,vcuvcko2021towards,sovrin2021principles} is the impossibility of tampering with communications between identity owners, \textit{i.e.} \emph{secure transactions}.

\emph{Delegation} is the final \emph{functional} characteristic of SSI proposed in the literature~\cite{sovrin2021principles}. It is the capacity of identity owners to delegate some of their identity data to other individuals or groups of individuals of their choosing. This is a developing field of study~\cite{Linklater2018b,Liu,lim2021subject}.

\subsubsection{Non-Functional}

According to the authors of~\cite{Andrieu2016,Naik2020a,VanWingerde2017a,Diebold2017a,vcuvcko2021towards}, a critical component of SSI is ensuring that people's data are \emph{recoverable} in the event of loss of personal device. This theoretical proposition is also an active area of applied research, as evidenced by six recent articles~\cite{Jakubeit2020,Soltani2019c,kostadinov2021towards,Linklater2018b,Kim2021a,singh2021private}.

Six studies assert that \emph{usability} is critical in SSI~\cite{Diebold2017a,Toth,Satybaldy2020d,Speelman2020a,vcuvcko2021towards,sovrin2021principles}. These works affirm that: (i) interfaces and experience must be optimized~\cite{vcuvcko2021towards,sovrin2021principles}; (ii) users' needs and expectations must be met and consistent across all platforms and services~\cite{Satybaldy2020d}; (iii) users should not require prior knowledge of blockchain technology~\cite{Diebold2017a}; as well as (iv) other underlying technologies such as cryptographic operations, biometrics, databases, and protocols~\cite{Toth}. One way to accomplish these goals is to mimic physical identities and the interactions we have with them, thereby exposing the user to familiar workflows~\cite{Toth}. Ultimately, if the user does not comprehend what is occurring and is unable to reason about it, the user is not sovereign~\cite{Speelman2020a}.

\emph{Accessibility} is a concept related to usability but has a more specific focus. According to three research papers in the reviewed literature, identity-related solutions should be accessible to as many people as possible~\cite{Naik2020a,sovrin2021principles,vcuvcko2021towards}.


Two authors claim that identities should always be \emph{available}~\cite{Naik2020a,Ferdous2019}. The challenge of having highly available identity-related information in SSI is being addressed on multiple fronts. For example, \cite{chotkan2021industry} and~\cite{abraham2020revocable} propose ensuring the availability of issuers' revocation registries in a decentralized and offline fashion.

In terms of \emph{auditability}, Schutte~\cite{Schutte2016} argued that auditing requires not only access to the details, but also the ability to read and understand them.

Another significant factor to consider is the \emph{scalability} of SSI systems~\cite{Naik2020a,Andrieu2016,Satybaldy2020d}. While practical research observes and considers this aspect~\cite{schardong2021matching,gruner2021atib,schanzenbach2020towards}, it is not the norm in the surveyed literature.

Finally, there is a subset of articles arguing for the importance of \emph{regulatory} compliance in the SSI ecosystem~\cite{VanWingerde2017a,chotkan2021industry}, such as the GDPR~\cite{gdpr2016} and CCPA~\cite{ccpa2018}. Chotkan \textit{et al.}~\cite{chotkan2021industry} argued for the importance of verification and legislation compliance, despite the fact that the latter may weaken the strength of other SSI principles (such as privacy). The author of~\cite{Abraham2017} did not say that GDPR compliance was necessary, but they discussed about how SSI systems can use verifiable claims to meet the following articles of the GDPR: (i) consent; (ii) pseudonymization; (iii) the right to be forgotten; (iv) records of processing activities; (v) data portability; and (vi) data protection by design and by default.

\subsection{Refute}

There are three works~\cite{Schutte2016,Toth,Ellingsen2019a} that add new properties to SSI while also refuting some of Allen's concepts~\cite{Path2016}. They all refute the \emph{existence} principle, which states that individuals cannot exist entirely in digital form, and that (self-sovereign) identities expose some aspects of the user. Toth and Anderson-Priddy~\cite{Toth} have also argued against \emph{transparency} and \emph{protection}, suggesting that more debate is needed on these topics. Similarly, the authors of~\cite{Ellingsen2019a} argued that previous discussions~\cite{Abraham2017,cameron2005laws} about identity had failed to address the issue of \emph{existence}.

Unlike the previous two studies, Schutte~\cite{Schutte2016} examined Allen's principles through a more philosophical and less technical lens. He contended that an individual, or ``self'' is not an indivisible entity, but rather the result of constant interactions between various agents, both internal and external. He then criticized the principles of \emph{existence}, \emph{control}, \emph{access}, and \emph{consent}, claiming that an individual's identity is a ``heuristic that simplifies information processing and decision making''~\cite{Schutte2016}, which is imprecise by nature and thus cannot fully anchor identity processes. Finally, he argued that claims are critical and can be viewed as signals broadcast by some actors and perceived by others, who must decide how to prioritize and interpret them.

\section{RQ-4: When, where, and by whom were SSI studies published?}\label{sec:RQ4}
To address \ref{rq-4}, we aggregate the \hyperref[general]{\emph{General}} data items gathered via our data extraction form. The following section discusses the findings.

\subsection{Frequency of publication}

\begin{figure}[H]
  \centering
  \input{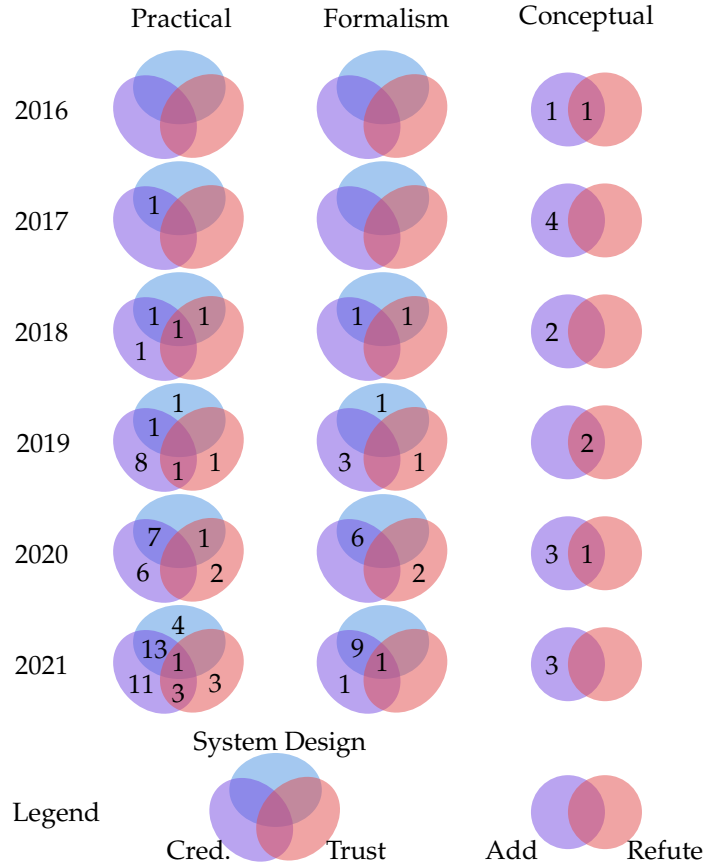}
  \caption{The number of publications in each facet of our taxonomy over time.}
  \label{fig:RY}
\end{figure}

In terms of publication frequency, Table~\ref{tbl:years} summarizes publications by year. Although it is a brief overview, it demonstrates the growing academic interest in SSI. Using Venn diagrams to represent the facets of our taxonomy, we can discern finer details regarding annual publication frequency. Figure~\ref{fig:RY} depicts the number of publications classified in this manner.

In response to Allen's introduction of the ten principles in 2016~\cite{Path2016}, two publications were released in the same year~\cite{Andrieu2016,Schutte2016}. Works published in 2016 and 2017 are mostly conceptual writings that expand on Allen's discussion, proposing new principles/requirements~\cite{Andrieu2016,Schutte2016,Abraham2017,Diebold2017a,VanWingerde2017a,VCfaq2017} for SSI as well as refuting some~\cite{Schutte2016}. Since 2016, researchers have been conducting continuous conceptual research, indicating that the meaning of SSI is still being debated. Beginning in 2018, articles started to significantly introduce new pragmatic problems and solutions to the SSI ecosystem, as well as mathematical formalisms. Nonetheless, mathematical formalization and formal description of cryptographic tools in applied research, which help SSI grow into a well-defined field of study, account for less than or equal to half of all applied research published each year.

\captionsetup[table]{singlelinecheck=on}

\begin{table}[H]
    \centering
    \caption{Publications per year.}
    \label{tbl:years}
    \begin{tabular}{lcp{31em}}
        \toprule
        \textbf{Year} & \textbf{Total} & \textbf{Studies} \\ \midrule
        2016 & 2 & \cite{Andrieu2016,Schutte2016} \\ \hdashline \addlinespace[1pt]
        2017 & 5 & \cite{Abraham2017,Diebold2017a,VanWingerde2017a,w3cVC,VCfaq2017} \\ \hdashline \addlinespace[1pt]
        2018 & 5 & \cite{Gruner2018,Stokkink2018c,Linklater2018b,Muhle,Schanzenbach2018a} \\ \hdashline \addlinespace[1pt]
        2019 & 14 & \cite{Gruner2019,Gruner2019a,Bathen2019,Ferdous2019,Toth,Soltani2019c,Lux2019,Ellingsen2019a,Lee2019,Schanzenbach2019,w3cDID,smith2019key,DIDExchangeProtocol,lagutin2019enabling} \\ \hdashline \addlinespace[1pt]
        2020 & 19 & \cite{Jakubeit2020,Satybaldy2020d,Abraham2020,Liu,Xu2020a,Hong2020a,Speelman2020a,Barclay2020a,Wohlgemuth2020,Naik2020a,inoue2020cooperative,Liu2020,Lux2020,Toth2020,bhattacharya2020enhancing,abraham2020revocable,schanzenbach2020towards,hardman2020didcomm,yang2020zero} \\ \hdashline \addlinespace[1pt]
        2021 & 37 & \cite{schardong2021matching,kubach2021lightweight,singh2021private,lim2021subject,lockwood2021accessible,lauinger2021poa,shanmugarasa2021towards,WO2021064182A1,WO2021125586A1,lax2021lightweight,abramson2021evaluating,Kim2021a,jaroucheh2021secretation,lemieux2021addressing,samir2021dt,gruner2021atib,lee2021privacy,kang2021decentralized,mishra2021pseudo,abraham2021ssi,martinez2021applying,yildiz2021connecting,siddiqui2021credentials,naik2021attack,cho2021verifiable,fedrecheski2021low,muktacredential,anaigoundanpudur2021cryptographic,chotkan2021industry,kostadinov2021towards,vcuvcko2021towards,park2021new,bobolz2021issuer,zhong2021jointcloud,sovrin2021principles,alber2021adapting,kim2021analysis} \\
        \bottomrule
    \end{tabular}
\end{table}

\subsection{Publishing Venues}

In terms of publication venues, forty-two papers were held in congresses, symposia, or forums, as shown in Table~\ref{tbl:venues} under the category conference. Forty-two conference publications and six master's theses show research momentum. However, it is still in its infancy, with just one Ph.D. thesis and fifteen journal articles.

\captionsetup[table]{singlelinecheck=off}

\begin{table}[H]
    \footnotesize
    \begin{adjustwidth}{-\extralength}{0cm}
    \caption{Types of publishing venues over the years.}
    \label{tbl:venues}
        \resizebox*{\linewidth}{!}{
        \begin{tabularx}{\linewidth}{rP{2em}P{3em}P{4em}P{4em}P{7.5em}P{9.5em}P{10.7em}}
            \toprule
            \textbf{Venue Type} & \textbf{Total} & \textbf{2016} & \textbf{2017} & \textbf{2018} & \textbf{2019} & \textbf{2020} & \textbf{2021} \\ \midrule
            Blog Post & 1 & \cite{Schutte2016} & & & & &  \\ \hdashline \addlinespace[1pt]
            Website & 2 & & \cite{VCfaq2017} & & & & \cite{sovrin2021principles} \\ \hdashline \addlinespace[1pt]
            Report & 1 & & \cite{Abraham2017} & & & & \\ \hdashline \addlinespace[1pt]
            Standard & 4 & & \cite{w3cVC} & & \cite{w3cDID,DIDExchangeProtocol} & \cite{hardman2020didcomm} &  \\ \hdashline \addlinespace[1pt]
            Web Archive & 7 & & & & \cite{Lee2019,Schanzenbach2019,smith2019key} & \cite{Barclay2020a} & \cite{vcuvcko2021towards,fedrecheski2021low,lee2021privacy} \\ \hdashline \addlinespace[1pt]
            Conference & 42 & \cite{Andrieu2016} & & \cite{Gruner2018,Stokkink2018c,Linklater2018b,Schanzenbach2018a} & \cite{Gruner2019,Gruner2019a,Bathen2019,Soltani2019c,Lux2019,lagutin2019enabling} & \cite{Jakubeit2020,Satybaldy2020d,Abraham2020,Wohlgemuth2020, Kim2021a,Naik2020a,Liu2020,Lux2020,Toth2020,bhattacharya2020enhancing,abraham2020revocable,inoue2020cooperative} & \cite{kubach2021lightweight,singh2021private,alber2021adapting,lim2021subject,kim2021analysis,lauinger2021poa,shanmugarasa2021towards,jaroucheh2021secretation,lemieux2021addressing,siddiqui2021credentials,yildiz2021connecting,naik2021attack,schardong2021matching,martinez2021applying,abraham2021ssi,muktacredential,zhong2021jointcloud,mishra2021pseudo,bobolz2021issuer}  \\ \hdashline \addlinespace[1pt]
            Journal & 15 & & & \cite{Muhle} & \cite{Ferdous2019,Toth} & \cite{Liu,Xu2020a,Hong2020a,yang2020zero} & \cite{lockwood2021accessible,lax2021lightweight,abramson2021evaluating,samir2021dt,gruner2021atib,cho2021verifiable,kang2021decentralized,park2021new}  \\ \hdashline \addlinespace[1pt]
            Patent & 2 & & & & & & \cite{WO2021125586A1,WO2021064182A1} \\ \hdashline \addlinespace[1pt]
            Bachelor Thesis & 1 & & & & & & \cite{kostadinov2021towards}  \\ \hdashline \addlinespace[1pt]
            Master Thesis & 6 & & \cite{Diebold2017a,VanWingerde2017a} & & \cite{Ellingsen2019a} & \cite{Speelman2020a} & \cite{anaigoundanpudur2021cryptographic,chotkan2021industry}  \\ \hdashline \addlinespace[1pt]
            PhD Thesis & 1 & & & & & \cite{schanzenbach2020towards} & \\
            \bottomrule
        \end{tabularx}}
	\end{adjustwidth}
\end{table}

Authors choose a wide variety of conferences, symposia, and forums in which to publish their works. Even though forty-two papers have been published in this sort of venue, only seven conferences have received more than one publication, as shown in Table~\ref{tbl:confs}. The IEEE colloquia, which received nineteen papers spread across fourteen different conferences, are the most popular choice. As illustrated in Table~\ref{tbl:journals}, the same trend holds true for essays published in scientific journals. Seven of the fifteen studies were published in journals published by the IEEE.

\begin{table}[h]
    \footnotesize
    \caption{Conferences, symposia and forums with multiple publications.}
    \label{tbl:confs}
    \begin{adjustwidth}{-\extralength}{0cm}
        \begin{tabularx}{\linewidth}{lCC}
            \toprule
            \textbf{Venue Name} & \textbf{Total} & \textbf{Studies} \\ \midrule
            Conference on Blockchain Research \& Applications for Innovative Networks and Services  & 2 & \cite{Liu2020,Lux2020} \\ \hdashline \addlinespace[1pt] 
            Open Identity Summit & 2 & \cite{kubach2021lightweight,alber2021adapting} \\ \hdashline \addlinespace[1pt]
            International Conference on Information Networking & 2 & \cite{lim2021subject,kim2021analysis} \\ \hdashline \addlinespace[1pt]
            IEEE Symposium Series on Computational Intelligence  & 2 & \cite{Gruner2019a,naik2021attack} \\ \hdashline \addlinespace[1pt]
            IEEE International Congress on Cybermatics & 2 & \cite{Gruner2018,Stokkink2018c}\\ \hdashline \addlinespace[1pt]
            IEEE International Conference on Blockchain and Cryptocurrency & 2 & \cite{lauinger2021poa,jaroucheh2021secretation} \\ \hdashline \addlinespace[1pt]
            IEEE International Conference on Trust, Security and Privacy in Computing and Communications & 2 & \cite{Schanzenbach2018a,abraham2020revocable} \\ \hdashline \addlinespace[1pt]
            IEEE International Conference on Internet of Things: Systems, Management and Security & 1 & \cite{Lux2019} \\ \hdashline \addlinespace[1pt]
            IEEE International Conference on Mobile Cloud Computing, Services, and Engineering & 1 & \cite{Naik2020a} \\ \hdashline \addlinespace[1pt]
            IEEE International Conference on Cloud Engineering & 1 & \cite{siddiqui2021credentials} \\ \hdashline \addlinespace[1pt]
            IEEE International Symposium on Network Computing and Applications & 1 & \cite{Gruner2019} \\ \hdashline \addlinespace[1pt]
            IEEE International Symposium on Dependable, Autonomic and Secure Computing & 1 & \cite{Soltani2019c} \\ \hdashline \addlinespace[1pt]
            IEEE International Conference on Pervasive Computing and Communications Workshops & 1 & \cite{shanmugarasa2021towards} \\ \hdashline \addlinespace[1pt]
            IEEE Annual Computers, Software, and Applications Conference & 1 & \cite{lemieux2021addressing} \\ \hdashline \addlinespace[1pt]
            IEEE Conference on Computer Vision and Pattern Recognition Workshops & 1 & \cite{Bathen2019} \\ \hdashline \addlinespace[1pt]
            IEEE International Performance, Computing, and Communications Conference & 1 & \cite{zhong2021jointcloud} \\ \hdashline \addlinespace[1pt]
            IEEE International Conference on Systems, Man, and Cybernetics & 1 & \cite{mishra2021pseudo} \\ \hdashline \addlinespace[1pt]
            IEEE Symposium on Computers and Communications & 1 & \cite{yildiz2021connecting} \\ \hdashline \addlinespace[1pt]
            IFIP International Conference on Information Security Theory and Practice & 1 & \cite{Jakubeit2020} \\ \hdashline \addlinespace[1pt]
            IFIP International Summer School on Privacy and Identity Management & 1 & \cite{Satybaldy2020d} \\ \hdashline \addlinespace[1pt]
            IFIP International Conference on New Technologies, Mobility and Security & 1 & \cite{singh2021private} \\ \hdashline \addlinespace[1pt]
            ACM Celebration of Women in Computing & 1 & \cite{muktacredential} \\ \hdashline \addlinespace[1pt]
            International Conference on Information and Communications Security & 1 & \cite{Abraham2020} \\ \hdashline \addlinespace[1pt]
            International Conference on Innovative Mobile and Internet Services in Ubiquitous Computing & 1 & \cite{Kim2021a} \\ \hdashline \addlinespace[1pt]
            International Conference on Security and Cryptography & 1 & \cite{abraham2021ssi} \\ \hdashline \addlinespace[1pt]
            International Teletraffic Congress & 1 & \cite{inoue2020cooperative} \\ \hdashline \addlinespace[1pt]
            International Symposium on Networks, Computers and Communications & 1 & \cite{bhattacharya2020enhancing} \\ \hdashline \addlinespace[1pt]
            International Conference on Business Process Management Workshops & 1 & \cite{schardong2021matching} \\ \hdashline \addlinespace[1pt]
            International Conference on Cryptology and Network Security & 1 & \cite{bobolz2021issuer} \\ \hdashline \addlinespace[1pt]
            Symposium on Cryptography and Information Security & 1 & \cite{Wohlgemuth2020} \\ \hdashline \addlinespace[1pt]
            Annual Privacy Forum & 1 & \cite{Toth2020} \\ \hdashline \addlinespace[1pt]
            Annual Conference of the South African Institute of Computer Scientists and Information Technologists & 1 & \cite{Linklater2018b} \\ \hdashline \addlinespace[1pt]
            Rebooting the Web-of-Trust & 1 & \cite{Andrieu2016} \\ \hdashline \addlinespace[1pt]
            Gesellschaft fur Informatik (GI) & 1 & \cite{martinez2021applying} \\ \hdashline \addlinespace[1pt]
            Workshop on Decentralized IoT Systems and Security & 1 & \cite{lagutin2019enabling} \\
            \bottomrule
        \end{tabularx}
	\end{adjustwidth}
\end{table}

\captionsetup[table]{singlelinecheck=on}

\begin{table}[h]
    \centering
    \caption{Studies published in journals.}
    \label{tbl:journals}
    \begin{tabular}{ l c c }
        \toprule
        \textbf{Journal Name} & \textbf{Total} & \textbf{Studies} \\ \midrule
        Frontiers in Blockchain & 3 & \cite{lockwood2021accessible,abramson2021evaluating,abramson2021evaluating} \\ \hdashline \addlinespace[1pt]
        IEEE Access & 2 & \cite{Ferdous2019,gruner2021atib} \\ \hdashline \addlinespace[1pt]
        IEEE Internet of Things Journal & 2 & \cite{samir2021dt,park2021new} \\ \hdashline \addlinespace[1pt]
        IEEE Software & 1 & \cite{Liu} \\ \hdashline \addlinespace[1pt]
        IEEE Security and Privacy & 1 & \cite{Toth} \\ \hdashline \addlinespace[1pt]
        IEEE Transactions on Vehicular Technology & 1 & \cite{Xu2020a} \\ \hdashline \addlinespace[1pt]
        IEEE Transactions on Computational Social Systems & 1 & \cite{lax2021lightweight} \\ \hdashline \addlinespace[1pt]
        Elsevier Computer Science Review & 1 & \cite{Muhle} \\ \hdashline \addlinespace[1pt]
        Elsevier Computers \& Security & 1 & \cite{yang2020zero} \\ \hdashline \addlinespace[1pt]
        MDPI Electronics & 1 & \cite{Hong2020a} \\ \hdashline \addlinespace[1pt]
        IEICE Transactions on Information and Systems & 1 & \cite{cho2021verifiable} \\ \hdashline \addlinespace[1pt]
        Ledger & 1 & \cite{kang2021decentralized} \\
        \bottomrule
    \end{tabular}
\end{table}

\subsection{Authors}\label{sec:authors}

\begin{figure}[h]
  \begin{adjustwidth}{-\extralength}{0cm}
      \input{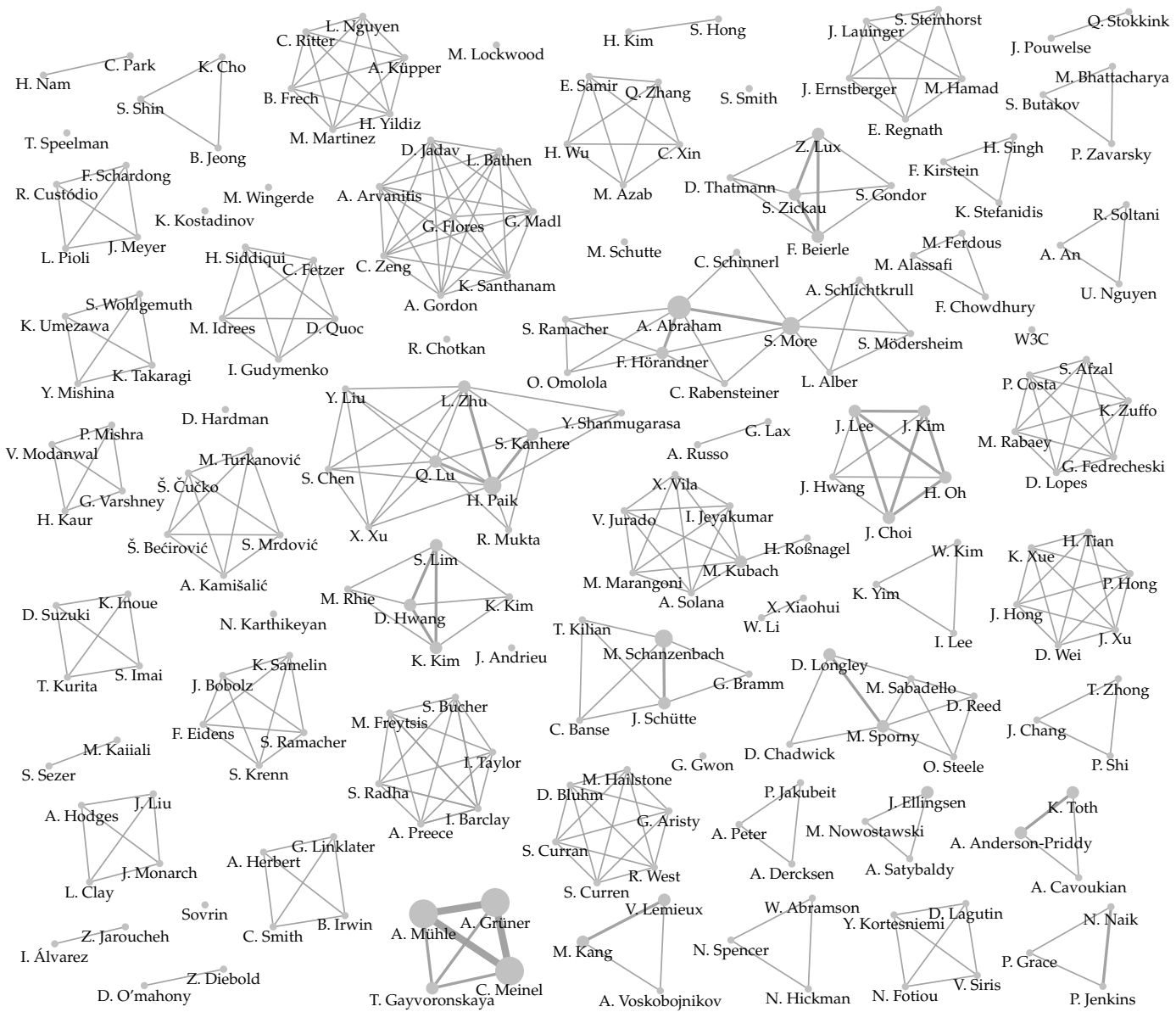}
      \caption{Co-authorship network graph, where vertices represent authors and edges their co-authorship of one or more works.}
      \label{fig:RA}
  \end{adjustwidth}
\end{figure}

We gathered the authors' names using our data extraction form. This allowed us to construct a co-authorship network graph~\cite{savic2019}, a weighted undirected graph in which vertices represent authors and edges represent works shared between them. Figure~\ref{fig:RA} depicts our co-authorship network graph, with edge weights displayed in different line diameters for ease of reading. The diameter of the vertices changes as well, representing the number of publications each author has. The vast majority of the edges in this network graph are thin, indicating that most authors only have one publication. Additionally, this disconnected graph shows that authors have mostly worked alone or in small groups. 

The authors with the most publications in this survey are Andreas Grüner, Alexander Mühle, and Christoph Meinel. They have co-authored three research papers~\cite{Gruner2019,Gruner2019a,gruner2021atib} and two more with Tatiana Gayvoronskaya~\cite{Gruner2018,Muhle}. As a result, the vertices and edges representing these three authors and their publications have the most weight in this graph (\textit{i.e.}, the thickest vertices and edges).

Andreas Abraham is the only author who has written four articles. Abraham's publications include a technical report~\cite{Abraham2017}, a research paper with Felix Hörandner, Olamide Omolola, and Sebastian Ramacher~\cite{Abraham2020}, a second paper with Felix Hörandner, Christof Rabensteiner, and Stefan More~\cite{abraham2020revocable}, and a third paper with the last two authors~\cite{abraham2021ssi}.

After introducing Andreas Abraham, who is a co-author of four publications, we now introduce the researchers who are co-authors of three: Stefan More, Martin Schanzenbach, and Hye-Young Paik. Apart from the two publications with Andreas Abraham, Stefan More also co-authored a research paper with Lukas Alber, Sebastian Mödersheim, and Anders Schlichtkrull~\cite{alber2021adapting}. Schanzenbach's publications include his doctoral dissertation~\cite{schanzenbach2020towards} and two articles co-written with Julian Schütte, one with Georg Bramm~\cite{Schanzenbach2018a}, and one with Thomas Kilian and Christian Banse~\cite{Schanzenbach2019}. Hye-Young Paik and Liming Zhu co-authored an article with Yue Liu, Qinghua Lu, Xiwei Xu, and Shiping Chen~\cite{Liu}, and Paik published another article with Yashothara Shanmugarasa and Salil S. Kanhere~\cite{shanmugarasa2021towards}. Paik also shares a third article with Rahma Mukta, Qinghua Lu, and Salil S. Kanhere~\cite{muktacredential}.

\begin{figure}[h]
  \begin{adjustwidth}{-\extralength}{0cm}
      \centering
      \input{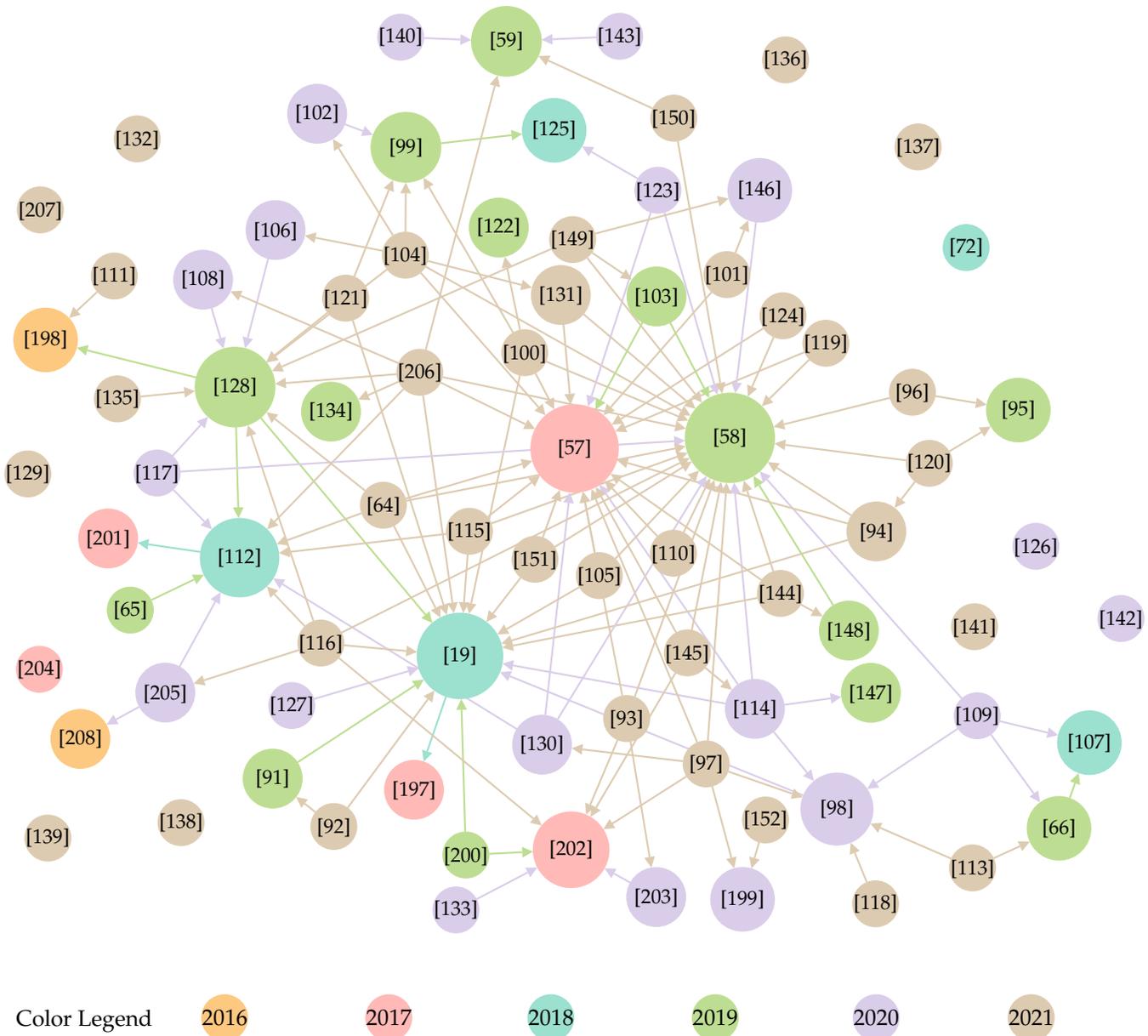}
      \caption{Co-reference network.}
      \label{fig:RP2}
  \end{adjustwidth}
\end{figure}

We present in Figure~\ref{fig:RP2} the co-reference network of the surveyed literature. The vertices in this directed graph represent publications. The edges represent references between articles, with the destination of an edge indicating that the source of the edge references this work. The number of received citations determines the diameter of the vertices, and the color of the vertices is determined by the year of publication.

This graph shows the significance of W3C standards DID~\cite{w3cDID} and VC~\cite{w3cVC} for SSI. They are the two most referenced works in this map, with twenty-nine and twenty-one references, respectively. The first survey of SSI~\cite{Muhle}, published in 2018, ranks third in terms of citations, with seventeen. It is followed by the fourth most cited article, a comprehensive mathematical formulation of SSI from 2019~\cite{Ferdous2019}.

In terms of cross-references, forty-seven works are not cited in any of the surveyed publications. Thirty-five of these unreferenced works are from 2021, nine from 2020, two from 2019, and one from 2018. Similarly, twenty-seven publications do not contain any references to mapped work. Eight of these are from 2021, three are from 2020, six are from 2019, three are from 2018, five are from 2017, and two are from 2016. The scope of our survey is one of the reasons for publications that do not include references to other mapped works. We excluded SSI platforms such as Sovrin, Uport, and Jolocom, which are mentioned in many of these essays.

\section{Open Challenges}\label{sec:challenges}
The surveyed materials detail developments in the field of SSI. New publications will advance the conceptual debate about what it means for an identity to be self-sovereign, while also introducing new and unexpected challenges to the SSI ecosystem. We identify future research challenges based on the evidence gathered to address our research questions. They are discussed in detail below, along with recommendations.

\emph{A definition of SSI} that researchers and practitioners accept. We have gathered evidence (see Section~\ref{sec:RQ3}) that the majority of articles on SSI fundamentals agree with Allen's principles~\cite{Path2016}, while also adding new ones. Promoting a thorough review and discussion is critical in order to develop a new set of rules for defining SSI. Furthermore, mathematical formalization can be used to define precise boundaries. Having an exact definition of SSI will benefit future efforts and, ultimately, users who will be able to transition between SSI systems with the confidence that they share the same fundamentals.

\emph{Fundamental research.} The majority of materials surveyed that include a mathematical model do so by designing it to their particular context. Only one of the articles reviewed provides a comprehensive mathematical formulation of SSI~\cite{Ferdous2019}, but it does not address the SSI's inherent decentralized trust properties. Another article~\cite{Liu} discusses realistic considerations and provides design patterns for numerous facets of SSI, including trust. These publications serve as a valuable starting point. However, additional basic research is necessary to foster discussion about how to jointly represent identities, credentials, claims, and trust, which is critical for future pragmatic research. By addressing \ref{rq-2} and \ref{rq-3} (see Sections~\ref{sec:RQ2} and \ref{sec:RQ3}), we established a foundation for future fundamental research.

\emph{Special case attribute sharing.} Revised publications allow VPs to: (i) selectively disclosure attributes~\cite{Abraham2020,abraham2020revocable,yang2020zero}; (ii) create Boolean predicates about attributes~\cite{bobolz2021issuer}; and (iii) produce general expressions over attributes~\cite{lee2021privacy,Lee2019,Schanzenbach2019}. Nonetheless, these methods are unsuitable when sharing characteristics that will likely stay unchanged for several years. For instance, the shipping address associated with an online purchase. As a result, additional research on VP is required to ensure that a diverse range of use cases is covered.

\emph{Sound trust models.} Trust plays an essential role in SSI and will be of paramount importance for the adoption of SSI solutions. Without comprehensive testing, trust models will become attractive targets for hackers. This open challenge is exacerbated by the current standardization effort~\cite{w3cTrustModel}, which specifies a Boolean trust model in which a verifier either trusts or distrusts the issuer. This model does not cover the fuzzy scenarios of the real world. For example, an entity may present multiple claims about the same attribute where some issuers are trusted and others are not. Can this claim be trusted? Quantifiable trust/reputation models are needed, but only five of the surveyed articles address this issue~\cite{Gruner2018,Gruner2019a,bhattacharya2020enhancing,zhong2021jointcloud,abramson2021evaluating}. Furthermore, trust models require strong security, so formal verification techniques must be employed~\cite{meier2013tamarin}.

\emph{Blockchainless SSI.} On blockchain-based SSI systems, dependence in centralizing authorities has been reduced but not eliminated entirely; instead, it has been replaced by a decentralized entity in which the user must place their trust in order to embrace SSI. To participate in an SSI ecosystem, the user should not be required to trust and rely on a blockchain consortium. However, the majority of publications operate under the erroneous assumption that blockchain is a necessary component of SSI. To be self-sovereign, the user should not have to trust anyone, not even a blockchain.

\emph{To facilitate the migration from other paradigms.} In federated and user-centric models, the IdP bears the administrative burden. Users need only to be concerned with their passwords. With SSI, users are also overburdened with management tasks such as backing up their keys, identities, and credentials, as well as creating and presenting claims. We mapped publications that propose techniques for deriving (self-sovereign) identities from federated and user-centric identities~\cite{Abraham2020,abraham2021ssi,Bathen2019}, as well as those that discuss backup and recovery~\cite{Soltani2019c,kostadinov2021towards,Linklater2018b,Jakubeit2020,Kim2021a,singh2021private}. As a result, academia is gaining momentum on this migration issue.

\emph{Usability}. Humans will interact with SSI systems. It is critical to research interfaces and how people engage with them, as well as how users interact with one another. Meaningful interaction must occur between users and applications and, more importantly, between individuals in an SSI ecosystem. Otherwise, users are unlikely to leave the comfort of their current federated/user-centric identities. A common trend in usability research in SSI is to mimic physical wallets~\cite{Toth2020,Speelman2020a}, thus presenting the user with everyday interactions. Innovative solutions are necessary and can be decisive for the widespread adoption and success of SSI.

\section{Final Remarks}\label{sec:conclusion}
SSI is a new and promising identity management paradigm that increases people's agency in the digital world. It is gaining popularity among academics and industry. We filled in the gaps left by existing surveys, which lack methodological rigor and present biased results in favor of blockchain, thus missing the bigger picture.

In this article, we systematically surveyed both peer-reviewed and non-peer-reviewed literature that: (i) expanded the conceptual discussion on what SSI is; (ii) used mathematical formulation to precisely define one or more SSI-related problems and what cryptographic and non-cryptographic tools were used to solve them; and (iii) introduced novel pragmatical problem related to the SSI ecosystem and present a solution to it. After keywording the selected materials, a novel taxonomy of SSI was proposed. 

To answer our four research questions, we conducted four separate investigations on the surveyed literature. The results were reported in accordance with the proposed taxonomy and summarized in tables. Maps and tables were also created to categorize the current state-of-the-art research in SSI. These resources, when combined, enable the reader to comprehend each contribution individually while also providing a broad understanding of the current state and maturity of research in SSI. The reported results of our systematic method serve as a foundation for researchers and entrepreneurs who wish to conceptually expand SSI or develop new SSI-related systems. Finally, we discussed unresolved issues and provided recommendations for future research.


\section*{Thanks go to the Federal Institute of Education, Science and Technology of Rio Grande do Sul (Instituto Federal de Educação, Ciência e Tecnologia do Rio Grande do Sul, IFRS), which allowed doctoral studies for Frederico Schardong.}

\bibliographystyle{unsrt} 
\bibliography{bibliography.bib}

\end{document}